# A framework for quantitative modeling and analysis of highly (re)configurable systems

Maurice H. ter Beek, Axel Legay, Alberto Lluch Lafuente and Andrea Vandin

**Abstract**—This paper presents our approach to the quantitative modeling and analysis of highly (re)configurable systems, such as software product lines. Different combinations of the optional features of such a system give rise to combinatorially many individual system variants. We use a formal modeling language that allows us to model systems with probabilistic behavior, possibly subject to quantitative feature constraints, and able to dynamically install, remove or replace features. More precisely, our models are defined in the probabilistic feature-oriented language QFLAN, a rich domain specific language (DSL) for systems with variability defined in terms of features. QFLAN specifications are automatically encoded in terms of a process algebra whose operational behavior interacts with a store of constraints, and hence allows to separate system configuration from system behavior. The resulting probabilistic configurations and behavior converge seamlessly in a semantics based on discrete-time Markov chains, thus enabling quantitative analysis. Our analysis is based on statistical model checking techniques, which allow us to scale to larger models with respect to precise probabilistic analysis techniques. The analyses we can conduct range from the likelihood of specific behavior to the expected average cost, in terms of feature attributes, of specific system variants. Our approach is supported by a novel Eclipse-based tool which includes state-of-the-art DSL utilities for QFLAN based on the Xtext framework as well as analysis plug-ins to seamlessly run statistical model checking analyses. We provide a number of case studies that have driven and validated the development of our framework.

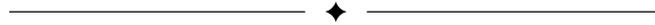

## 1 INTRODUCTION

SOFTWARE Product Line Engineering (SPLE) is a software engineering methodology aimed at developing, in a cost-effective and time-efficient manner, a family of software-intensive highly configurable systems by systematic reuse [1], [2], [3]. Individual products (or variants) share an overall reference variability model of the family, but they differ with respect to specific features, which are (user-visible) increments in functionality. The explicit introduction and management of feature-based variability in the software development cycle causes complexity in the modeling and analysis of software product lines (SPLs). There is a lot of recent research on lifting successful high-level algebraic modeling languages and formal verification techniques known from single (software) system engineering, such as process calculi and model checking, to SPLE (cf., e.g., [4], [5], [6], [7], [8], [9], [10], [11], [12], [13]). The challenge is to handle the variability inherent to SPLs, and to highly configurable systems in general, by which the number of possible variants may be exponential in the number of features. This paper presents our approach to this challenge.

**Modeling with the FLAN family**

We have developed a modeling approach based on a family of process-algebraic specification languages (FLAN [14], PFLAN [15] and QFLAN [16], surveyed in the invited contribution [17]). This family is inspired by the concurrent constraint programming paradigm of [18], its adoption in process calculi [19], and its stochastic extension [20].

In [14], we introduced the feature-oriented language FLAN. In FLAN, a rich set of process-algebraic operators allows one to specify both the configuration (in terms of features) and the behavior (in terms of actions) of products, while a constraint store allows one to specify all common constraints on features known from variability models such as feature models, as well as additional feature-based constraints on actions. The execution of a process is constrained by the store (e.g. to avoid introducing inconsistencies), but a process can also query the store (e.g. to resolve configuration options) or update the store (e.g. to add new features, even at runtime). The main distinguishing modeling feature of FLAN is a clean separation between the configuration and runtime aspects of an SPL.

In [15], we subsequently equipped FLAN with the means to specify probabilistic models of SPLs, resulting in PFLAN. PFLAN adds to FLAN the possibility to equip each action (including those that install an additional feature, possibly at runtime) with a rate, which can represent notions like uncertainty, failure rates, randomisation or preferences. This allows one to model and analyze the likelihood of installing features, the probabilistic behavior of users of products of the SPL and the expected average cost of products, next to probabilistic quantifications of ordinary temporal logic properties.

A fact that emerged during our experimentation with PFLAN was the need to consider a number of further aspects in the specification and analysis of behavioral models of SPLs, such as the staged configurations known from dynamic SPLs [21], [22] (e.g. adding and removing features as well as activating and deactivating features) and rich quantitative constraints (e.g. pricing constraints) over feature attributes reminiscent of [23].

For this purpose, we proposed the feature-oriented language QFLAN [16] as an evolution of probabilistic PFLAN [15]. QFLAN enriches PFLAN with the possibility to not only install but also uninstall and replace features at runtime as well as with advanced quantitative constraint modeling options regarding the 'cost' of features, i.e. feature attributes related to non-functional aspects such as price, weight, reliability, etc. In particular, the novel modeling op-tions we introduced were (i) quantitative constraints in the form of arithmetic relations among feature attributes (e.g. the total cost of a set of features must be less than a certain threshold); (ii) propositions relating the absence or presence of a feature to such a quantitative constraint (e.g. if a certain feature is present, then the total cost of a set of features must be less than a certain threshold); and (iii) rich action constraints involving such quantitative constraints (e.g. a certain action can be performed only if the total cost of the set of features constituting the product is less than a certain threshold). The uninstallation and replacement of features can be the result of malfunctioning or of the need to install a better version of the feature (e.g. a software update). We will illustrate this in our case studies, as well as the use of each of the above type of quantitative constraints over feature attributes, by providing concrete examples. It is important to note that the above type of quantitative constraints are significantly more complex than the ones that are commonly associated to attributed feature models [23], [24], [25], [26]. We are not aware of any other approach dealing with all of the above aspects in one unifying framework.

**Analysis tools for the FLAN family**
Our first tool support for the FLAN family was a prototypical implementation of an interpreter in MAUDE [27], which allowed us to conduct analyses on FLAN models, ranging from consistency checking (by means of SAT solving) to model checking. The introduction of PFLAN to model probabilistic aspects led us to develop corresponding tool support. We combined our MAUDE interpreter with the distributed statistical model checker MULTIVESTA [28], which allowed us to estimate the likelihood of specific system configurations and behavior, and thus to measure non-functional aspects such as quality of service, reliability or performance.

When QFLAN was eventually introduced, it became evident that, as feature attributes were typically not Boolean [23], the problem of deciding whether or not a product satisfies an attributed feature model with quantitative constraints, required more general satisfiability-checking techniques than SAT solving. This naturally led us to the use of Satisfiability Modulo Theory (SMT) solvers like Microsoft's Z3 [29], which allowed us to deal with richer notions of constraints like arithmetic ones. In fact, an important contribution of [16] was the integration of SMT solving into our approach, by means of a combination of our MAUDE QFLAN interpreter and Z3. In this paper, we present a complete re-engineering of the tool, in which we reimplemented from scratch our QFLAN simulator using Java. First, since we shifted the focus to quantitative analysis of fully configured (but variable) single products we did not need the full power of constraint *solving* anymore but just the simple constraint *checking*. We could hence replace the costly interaction with Z3 with an ad-hoc constraint evaluator for QFLAN using Java. Moreover, since the language was stable, we replaced the (slower) Maude executable semantics by an ad-hoc (faster) Java one. This re-engineering resulted in a mature tool with a modern integrated development environment for QFLAN.

Formally, our statistical model checking approach consists of performing a sufficient (and minimal) number of probabilistic simulations of a system model to obtain statistical evidence (with a predefined level of statistical confidence) of the quantitative properties being verified. Such properties are formulated in MULTIVESTA's property specification language MultiQuaTEx [28]. Statistical model checking offers unique advantages over exhaustive (probabilistic) model checking. First, statistical model checking does not need to generate entire state spaces and hence scales better without suffering from the combinatorial state-space explosion problem typical of model checking. In particular in the context of highly configurable systems, given their possibly combinatorially many variants, this outweighs the main disadvantage of having to give up on obtaining exact results (100% confidence) with exact analysis techniques like (probabilistic) model checking. Second, statistical model checking scales better with hardware resources, since the set of simulations to be carried out can be trivially parallelized and distributed. MULTIVESTA, indeed, can be run on multi-core machines, clusters or distributed computers with almost linear speedup. A further unique advantage of MULTIVESTA is that it can use the same set of simulations for checking several properties at the same time, thus offering even further reductions of computing time.

To the best of our knowledge, we were the first to apply statistical model checking in SPLE in [15]. There are other approaches to probabilistic model checking of SPLs [30], [31], [32], [33], [34], of which the latter comes closest to ours. In [34], the PROFEAT tool is presented. It provides a guarded-command language for modeling families of probabilistic systems and an automatic translation of such SPL models to the (featureless) input language of the probabilistic model checker PRISM [35]. It caters for the activation and deactivation of features at runtime and quantitative constraints over feature attributes. We will come back to this and other related work in Section 8.

**Contribution**
This paper provides a comprehensive presentation of our approach to the quantitative modeling and analysis of dynamic SPLs and other highly (re)configurable systems. With respect to our previous work on the FLAN family of modeling languages and its tool support, QFLAN has been extended and presented as a DSL with advanced Eclipse-based tool support. New higher-level languages have been incorporated to describe system behavior and property specifications. In particular, the designer can now decide to use either the process-algebraic language introduced in previous papers or a declarative rule-based language in the style of guarded command languages. This new process specification language in itself does not increase the expressive power of QFLAN, but it eases modeling as it is more amenable for visual representations as state machines.

In addition, the language has been extended with variables that can be queried and updated at runtime. Variables increase the expressive power of QFLAN and simplify the modeling of state-based information. The novel tool support eases the modeling and analysis task by providing an Eclipse editor for QFLAN specifications with integrated plug-ins for the analysis. The editor's features include auto-completion, error and syntax highlighting, as well as checks on the consistency of the specification and the initial models. In particular, the syntax guides the construction of consistent feature diagrams, and checks are done on initial models before starting their analysis. As mentioned before, several modules of our tool have been re-implemented. In particular, we have replaced two external back-end modules (the Z3 solver and the Maude interpreter) by internal modules. This provides savings by avoiding unnecessary data conversions. More significantly, our ad-hoc Java-based interpreter is faster since Maude does not have efficient libraries for certain data structures, and our constraint evaluator solves a more specific problem (constraint checking) than what we were exploiting from Z3 (constraint solving).

Last but not least, we validate our approach using several case studies. One of them was used in previous work and contributed to shape our approach, while two others show the scalability and versatility of our approach.

**Structure of the paper**

The paper outline is as follows. Section 2 presents a running example from [16] that we use throughout the paper to illustrate the main concepts of our approach. Section 3 introduces the high-level DSL we developed for QFLAN, followed by a presentation of the dynamics of the case study in Section 4. QFLAN's Eclipse-based tool support is presented in Section 5. Statistical analysis of QFLAN models with MULTIVESTA is introduced in Section 6, applied to the running example, followed by experimental quantitative analyses of two further case studies in Section 7. Section 8 discusses related work. Section 9 summarizes our contributions and lists possible future work.

## 2 RUNNING EXAMPLE: BIKES PRODUCT LINE

We introduce here a case study from [16] that we use as a running example to illustrate the main concepts of our approach and to provide intuitive cases of its possibilities and limitations. The case study stems from a collaboration with Bicincittà S.r.l. (www.bicincitta.com) and PisaMo S.p.A. (www.pisamo.it) in the context of the European project QUANTICOL (www.quanticol.eu). PisaMo is an in-house public mobility company of the Municipality of Pisa responsible for the introduction of the public bike-sharing system *CicloPi* in the city of Pisa two years ago. This bike-sharing system is supplied and monitored by Bicincittà.

To create an attributed feature model of a product line of bikes, we performed requirements elicitation on a set of documents generously shared with us by Bicincittà. This allowed us to extract the main features of the bikes they sell as part of the bike-sharing system, including indicative prices, and to identify their commonalities and variabilities. We then added some features that we found by reading through a number of documents on the technical characteristics and prices of bikes and their components as currently being sold by major bike vendors. The resulting model, which will be presented in the next section, thus has more variability than typical in bike-sharing systems. Indeed, vendors of such systems traditionally allow little variation to their customers (e.g. most vendors only sell bikes with a so-called step-thru frame, a.k.a. open frame or low-step frame, typical of utility bikes instead of considering other kinds of frames as we do), in part due to the difficulties of analyzing systems with high variability to provide guarantees on the deployed products and services. We believe that the progress of SPL analysis techniques (including the contribution of this paper) will help the adoption and hence the provision of richer (bike-sharing) systems with more variability. This is confirmed by the feedback we received during recent meetings with representatives of Bicincittà and PisaMo.

## 3 MODELING WITH QFLAN

The feature-oriented language QFLAN is the most recent member of the FLAN family (FLAN [14], PFLAN [15], QFLAN [16]) of process algebras inspired by the concurrent constraint programming paradigm of [18], its adoption in process calculi [19], and its stochastic extension [20]. QFLAN separates declarative (pre-)configuration aspects from procedural runtime aspects. It does so by using constraint stores, which allow the modeler to specify all common constraints from feature models (and more) in a *declarative* manner, while a rich set of process-algebraic operators allows the modeler to specify the configuration and behavior of product lines in a *procedural* manner. The semantics unifies *static* (pre-configuration) and *dynamic* (runtime) feature selection/installation.

In order to make our framework accessible by practitioners, we present QFLAN as a high-level DSL for which we offer a modern Eclipse-based integrated development environment. Moreover, we provide a detailed and rigorous presentation of the syntax and semantics of QFLAN to help practitioners univocally understand the intended semantics of specifications. In the rest of the paper, we freely use QFLAN to refer to both the presented DSL and the formal language. We first describe QFLAN specifications and their components (Sections 3.1–3.7) and then QFLAN initial models and their operational semantics (Section 3.8).

### 3.1 QFLAN specifications

Formally, a QFLAN specification is defined as follows.

**Definition 1 (QFLan specification).** A QFLAN specification $\mathcal{S}$ is a tuple $\langle \mathcal{F}, \mathcal{P}, \mathcal{V}, \mathcal{A}, \mathcal{C}, \mathcal{B} \rangle$ where:

- $\mathcal{F}$ is a finite set of features;
- $\mathcal{P}$ is a finite set of predicates;
- $\mathcal{V}$ is a finite set of variables;
- $\mathcal{A}$ is a finite set of actions;
- $\mathcal{C}$ is a finite set of constraints;
- $\mathcal{B}$ is a finite set of behavior.

For the rest of the paper, we fix a specification $\mathcal{S}$ as a tuple $\langle \mathcal{F}, \mathcal{P}, \mathcal{V}, \mathcal{A}, \mathcal{C}, \mathcal{B} \rangle$ for the sake of readability.

Each component of a QFLAN specification is specified in the DSL in one or several definition blocks as summarized in Table 1, which also contains a reference to the section where the components are described.

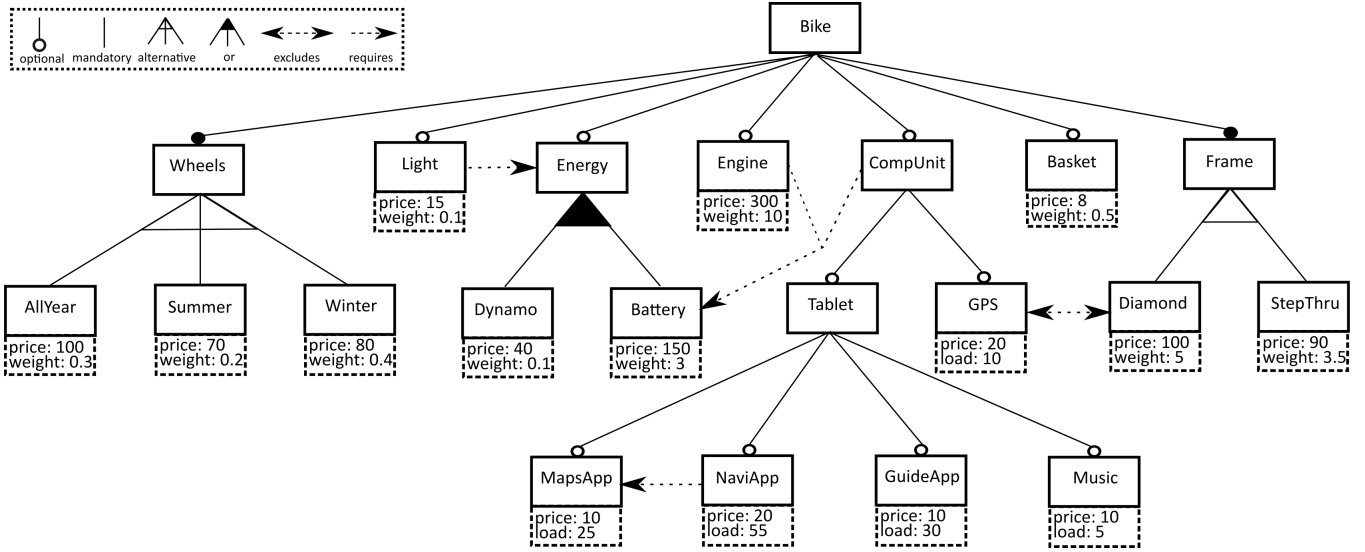

Fig. 1. Attributed feature model of bikes product line

TABLE 1
QFLᴀɴ specification components

| Component | Block | Section |
|---|---|---|
| Features | `abstract features`<br>`concrete features` | Section 3.2 |
| Predicates | `feature predicates` | Section 3.3 |
| Variables | `variables` | Section 3.4 |
| Actions | `actions` | Section 3.5 |
| Constraints | `feature diagram`<br>`cross-tree constraints`<br>`quantitative constraints`<br>`action constraints` | Section 3.6 |
| Behavior | `processes diagram`<br>`process` | Section 3.7 |

### 3.2 Features

Features in $\mathcal{F}$ represent user-visible properties or capabilities of products, over which a product line is defined. As in object-oriented programming, it is sometimes convenient to organize features and relate them to each other, possibly in hierarchies, resulting in so-called *feature diagrams*. For the bike product line of our running example, features such as the engine, the basket, the light, etc., are organized in the feature diagram depicted in Fig. 1.

More details on feature diagrams, including their contribution to the set $\mathcal{C}$ of constraints of a QFLᴀɴ specification, will be presented in the next section. For now, it is sufficient to know that they allow us to distinguish between two disjoint sets: *abstract* features $\mathcal{F}_A$ and concrete features $\mathcal{F}_C$. For instance, in our running example the wheel will be considered as an abstract feature, with different kinds of wheels (summer, winter, etc.) as concrete features. The complete set of all features of our running example is specified in QFLᴀɴ as shown in Listing 1. The declaration of abstract and concrete features is done in separate blocks `abstract features` and `concrete features`, respectively, and consists of a simple enumeration of the features. The set $\mathcal{F}$ of all features in a QFLᴀɴ specification is the union of such abstract and concrete features, i.e. $\mathcal{F} = \mathcal{F}_A \cup \mathcal{F}_C$.

```
begin abstract features                          1
  Bike Wheels Energy CompUnit Frame Tablet       2
end abstract features                            3
                                                 4
begin concrete features                          5
  AllYear Summer Winter Light Dynamo Battery Engine   6
  MapsApp NaviApp GuideApp Music GPS Basket Diamond   7
  StepThru Trashed                               8
end concrete features                            9
```

Listing 1. Features of the running example

A *product* is uniquely characterized by a non-empty subset of $\mathcal{F}$, its *installed* features, while a *product family* is characterized by a set of subsets of $\mathcal{F}$ (i.e. a set of products). It is worth noticing the product explosion typical of SPLs. For instance, in our running example the 21 features yield 1,314 different products. As we shall see later, products are subject to constraints such that not all feature combinations yield valid products. Indeed, constraints can partially reduce the number of products (e.g. some bikes may be too expensive, or too heavy, etc.) but not necessarily so much as to mitigate the inherent exponential explosion.

### 3.3 Feature predicates

QFLᴀɴ allows the modeler to consider static attributed features.

A QFLᴀɴ specification includes a finite set $\mathcal{P}$ of feature predicates, which are mappings $\mathcal{F} \to \mathbb{R}$ from $\mathcal{F}$ into real numbers (implemented as floats in the DSL). Only the mapping of concrete values in $\mathcal{F}_C$ can be explicitly specified in the DSL. Attribute values of abstract features in $\mathcal{F}_A$ cannot explicitly be specified and are instead automatically calculated as the sum of the predicates of their descendant concrete features which are currently installed. This will become more clear in the presentation of feature diagrams in Section 3.6.

Each concrete feature can indeed be equipped with a set of non-functional attributes. In our running example, for instance, we consider the attributes `price`, `weight` and

`load`, which respectively represent the specific feature's price in euros, weight in kilos, and computational load in percentage (of system utilization). The complete specification of attributes can be found in Listing 2. The name of the declaration block is `feature predicates` and it includes, for every attribute, the list of concrete features that have the attribute, together with the value of the attribute (if unspecified, by default concrete features have value zero for that predicate).

```
1  begin feature predicates
2    price = { AllYear = 100 , Summer = 70 , Winter = 80 ,
3            Light = 15 , Dynamo = 40 , Battery = 150 ,
4            Engine = 300 , MapsApp = 10 , NaviApp = 20 ,
5            GuideApp = 10 , Music = 10 , GPS = 20 ,
6            Basket = 8 , Diamond = 100 , StepThru = 90 }
7    weight = { AllYear = 0.3 , Summer = 0.2 , Winter = 0.4 ,
8            Light = 0.1 , Dynamo = 0.1 , Battery = 3 ,
9            Engine = 10 , Basket = 0.5 , Diamond = 5 ,
10           StepThru = 3.5 }
11   load = { MapsApp = 25 , NaviApp = 55 , GuideApp = 30 ,
12           Music = 5 , GPS = 10 }
13 end feature predicates
```
Listing 2. Attributes of the running example

### 3.4 Variables

QFLAN specifications include a set of real-valued variables $\mathcal{V}$, whose values can change during execution of a model. This is a novelty with respect to previous versions of the language, introduced to simplify modeling. Variables allow, e.g., to encode state information or context information (e.g. to model context-aware SPLs [25], [26]).

In our running example, variables are used in the analysis phase (e.g. to study properties of bikes at first deploy), or in constraints (not shown in our running example, but, e.g., used in the model in Section 7.4).

As shown in Listing 3, variables are specified in the `variables` block. The block also specifies the initial values of the variables, which contribute to defining the initial model, as explained later in Section 3.8.

```
1  begin variables
2    deploys = 0
3    trashed = 0
4  end variables
```
Listing 3. Variables of the running example

### 3.5 Actions

As discussed, QFLAN specifications consist of a declarative part, and of a procedural (or behavioral) part that will be described later. A key aspect of behavior are *actions*, which represent atomic runtime operations of a product.

We distinguish between *feature actions*, *user-defined actions* and *store actions*.

**Definition 2 (actions).** The set of actions $\mathcal{A}$ of a QFLAN specification $\mathcal{S}$ is the union of:

- the set of features $\mathcal{F}$, i.e. each installed feature can be used as an action;
- a finite set of user-defined actions $\mathcal{A}_u$;
- the set $\mathcal{A}_s$ of store actions composed by `install`($f$), `uninstall`($f$), `replace`($f, g$), and `ask`($C$), where $f, g \in \mathcal{F}$ are features and $C$ is a constraint (defined later in Section 3.6).

For each feature $f$, an action $f$ is implicitly included in the specification as a feature action, which represents the activation of the feature's main functionality (e.g., execution of `Music` models the fact that the biker turned on the music).

The set $A$ of user-defined actions is specified in a block `action`. In our running example this is shown in Listing 4, which contains for instance the actions `sell` and `dump` modelling, respectively, the selling and dumping of a bike.

```
1  begin actions
2    sell dump maintain book stop break start assistance
3  deploy
4  end actions
```
Listing 4. Additional actions of the running example

Store actions instead consist of: `install`($f$) (dynamic installation of a feature $f$), `uninstall`($f$) (dynamic uninstallation of a feature $f$), `replace`($f, g$) (dynamic replacement of feature $f$ by $g$) and `ask`($C$) (to query the store for the validity of constraint $C$).

### 3.6 Constraints

The declarative part of QFLAN is represented by a store of constraints $\mathcal{C}$, which impose relations and conditions on features, attributes and behavior. In particular, the QFLAN tool allows the modeler to specify the following classes of, conceptually different, constraints: *hierarchical constraints* $\mathcal{C}_H$, *cross-tree constraints* $\mathcal{C}_T$, *quantitative constraints* $\mathcal{C}_Q$, and *action constraints* $\mathcal{C}_A$. Our DSL provides convenient syntactic constructs to specify each of these classes of constraints in dedicated blocks with ad-hoc concrete syntax.

In our formal models, we represent all constraints uniformly in a constraint store $C$, which extends the specification constraints $\mathcal{C}$ with additional information as we shall see later in Section 3.8. In our formal notation, the union of two constraint stores $\mathcal{C}_1, \mathcal{C}_2$ is denoted by $\mathcal{C}_1 \oplus \mathcal{C}_2$. In particular, $\mathcal{C} = \mathcal{C}_H \oplus \mathcal{C}_T \oplus \mathcal{C}_Q \oplus \mathcal{C}_A$. For the formal semantics and its implementation in the tool, one key important aspect is that constraints come equipped with a notion of *consistency* and a notion of *entailment*. *consistent*($C$) amounts to logical satisfiability of all constraints constituting $C$, i.e. all constraints in $C$ are compatible with each other. Entailment of constraint $C'$ in $C$, denoted by $C \vdash C'$, amounts to logical entailment, i.e. $C'$ can be derived from $C$. Actually, the tool uses a simplified version of those problems. This is further formalized and explained at the end of the section and in Section 5.

#### Hierarchical constraints

The standard approach to express feature constraints in the SPL community is by means of a *variability model* which structures the features in the aforementioned *feature diagrams*, possibly enriched with cross-tree constraints and attributes. Such diagrams provide a convenient visual notation for specifying valid feature combinations. As discussed, the variability model of our bikes example is given in Fig. 1.

In QFLAN such a variability model is specified by providing the feature diagram and its additional cross-tree constraints in separate blocks. For example, the tree-like structure of our running example can be found in Listing 5. The `feature diagram` block has an enumeration of hierarchical relations of the form $f \otimes F \uplus F'$ where $F \subseteq \mathcal{F}$ is a set of *mandatory* features, $F' \subseteq \mathcal{F}$ is a set of *optional* features, and $\otimes \in \{\text{->}, \text{-OR->}, \text{-XOR->}\}$ is a father-child relation. In particular, $f$ is the father feature node in the tree and features in $F \uplus F'$ are the direct descendant child feature nodes. Features appearing as leaves must be concrete features, while features corresponding to internal nodes must be abstract features.

```
1  begin feature diagram
2    Bike -> { Wheels , ?Light , ?Energy , ?Engine ,
3              ?CompUnit , ?Basket , Frame }
4    Wheels -XOR-> { AllYear , Summer , Winter }
5    Energy -OR-> { Dynamo , Battery }
6    CompUnit -> { ?Tablet , ?GPS }
7    Frame -XOR-> { Diamond , StepThru }
8    Tablet -> { ?MapsApp , ?NaviApp , ?GuideApp , ?Music }
9  end feature diagram
```

Listing 5. Hierarchical constraints of the running example

The set of hierarchical constraints $\mathcal{C}_H$ is the composition of the constraints obtained from each father-child relation $h$, i.e. if we call $H$ the set of all such relations, then $\mathcal{C}_H = \bigoplus_{h \in H} \llbracket h \rrbracket$, where $\llbracket h \rrbracket$ formalizes the semantics of $h$ as a logical constraint on features.

**Definition 3 (semantics hierarchical relations).** Let $F$ be a set of mandatory features and let $F'$ be a set of optional features. The constraints of a hierarchical relation are defined by:

$$\llbracket f\text{->}F \uplus F' \rrbracket \equiv \bigwedge_{g \in F} \text{has}(g)$$
$$\llbracket f\text{-OR->}F' \rrbracket \equiv \text{has}(f) \to \bigvee_{g \in F'} \text{has}(g)$$
$$\llbracket f\text{-XOR->}F' \rrbracket \equiv \text{has}(f) \to \underline{\bigvee}_{g \in F'} \text{has}(g),$$

where $\text{has}(f)$ is a predicate denoting the presence of $f$ and $\underline{\vee}$ stands for the *exclusive or* logical operation.

The three kinds of relations $\otimes$ correspond to the three kinds of edges appearing in Fig. 1. The *and* relation $\text{->}$, used, e.g., to relate `Bike` with `Wheels-Frame`, enforces that all descendant features must be present in any product. This constraint can be relaxed by prefixing descendant features with ? (denoted with a circle in Fig. 1, as is common in feature diagrams), to indicate that the specific feature is optional, meaning that it may be present in a product. Descendant features not prefixed by ? are said to be mandatory (black dots). Hence, the relation from `Bike` to `Wheels-Frame` imposes any bike to have `Wheels` and `Frame`, while the presence of `Light`, `Energy`, `Engine`, `CompUnit` and `Basket` is optional. The *or* relation $\text{-OR->}$, used, e.g., to relate `Energy` with `Dynamo` and `Battery`, enforces that at least one descendant feature must be present in any product. The *xor* relation $\text{-XOR->}$, used, e.g., to relate `Wheels` with `AllYear`, `Summer` and `Winter`, enforces that precisely one descendant feature must be present in any product. All this, as well as other well-formedness conditions, is automatically checked by our tool to ease the task of the modeler.

We remark that the ? prefix can be used only for $\text{->}$ relations, since all child features of $\text{-OR->}$ and $\text{-XOR->}$ relations are by definition optional. Indeed, this is exactly as in FeatureIDE [36], [37], one of the most widely used tools for the modeling and analysis of classical feature models (i.e. devoid of numeric attributes and the likes). Moreover, all descendants of optional features must either be reached via $\text{-OR->}$ or $\text{-XOR->}$ relations, or be marked as optional. Note that this excludes the existence of a mandatory child feature of an optional feature. This is an implementation choice in our DSL, namely we consider the fact that a feature is mandatory to be a global declaration: mandatory features are by definition present in all configurations. This is a difference with FeatureIDE, where the fact that a feature is mandatory only models the dependency to its parent (i.e. even if a feature is marked as mandatory, it may be omitted in a configuration if the parent feature is omitted as well).

Also, we assume that only concrete features can be explicitly installed, while abstract features are implicitly installed as soon as one of its descendant features is installed.

This is formalized by assuming that any set of constraints $C$ is implicitly closed under the following axiom:

$$\frac{C \supseteq \text{has}(g) \;\; (f \otimes (g \cup F)) \in H}{C \supseteq \text{has}(g) \oplus \text{has}(f)}$$

The QFLAN interpreter automatically applies such closure.

We are now in a position to explain how predicates are evaluated for abstract features: they are computed as the sum of the predicate value of the installed concrete features which descend from the abstract feature.

In particular, given a set of constraints $C$, the set of father-child relations $H$, a predicate $p \in \mathcal{P}$, and an abstract feature $f \in \mathcal{F}_H$ the value of $p(f)$ is recursively defined by:

$$p(f) = \sum_{(f \otimes F) \in H, \, g \in F, \, C \vdash \text{has}(g)} p(g)$$

This is very useful, as we can for instance easily refer to the price of an entire bike with `price(Bike)`, or to the computational load of a tablet with `load(Tablet)`.

#### Cross-tree constraints

Features can also be related with cross-tree constraints of the form '$f$ requires $g$' and '$f$ excludes $g$'. The set of cross-tree constraints $\mathcal{C}_T$ is the composition of the constraints obtained from each cross-tree constraint $t$, i.e. if we call $T$ the set of all such relations, then $\mathcal{C}_T = \bigoplus_{t \in T} \llbracket t \rrbracket$, where $\llbracket t \rrbracket$ formalizes the semantics of $t$ as a logical constraint on features.

**Definition 4 (semantics cross-tree constraints).** Let $f$ and $g$ be features. The semantics of $f$ requires $g$ and $f$ excludes $g$ is defined as follows:

$$\llbracket f \text{ requires } g \rrbracket \equiv \text{has}(f) \to \text{has}(g)$$
$$\llbracket f \text{ excludes } g \rrbracket \equiv \text{has}(f) \to \neg\text{has}(g)$$

The cross-tree constraints of our running example can be found in Listing 6, enumerated under the block `cross-tree constraints`. It contains two common cross-tree relations. The relation of the form $f$ requires $g$ indicates that whenever feature $f$ (a node in the tree) is installed in a product, then also feature (node) $g$ must be installed, whereas $f$ excludes $g$ indicates that features $f$ and $g$ cannot be both present in the same product.

```
1  begin cross-tree constraints
2    Light requires Energy
3    Engine requires Battery
4    CompUnit requires Battery
5    NaviApp requires MapsApp
6    GPS excludes Diamond
7  end cross-tree constraints
```

Listing 6. Cross-tree constraints of the running example

*Quantitative constraints*

QFLAN also allows to specify quantitative constraints based on arithmetic relations among feature attributes. Formally, quantitative constraints in $\mathcal{C}_Q$ are Boolean formulas whose atomic propositions are numerical relations ($<$, $\leq$, $=$ and so on) between arithmetic expressions of real values, real-valued feature predicates (attributes), and variables. For our running example, we consider the following constraints:

(C1) a bike may cost at most 600 euros;
(C2) a bike may weigh up to 15 kilograms;
(C3) a bike's computational load may not exceed 100%.

Constraints (C1)–(C3) are part of the constraint store of our QFLAN model of the bikes product line. As such, they prohibit the execution of any action (e.g. the runtime (un)installation or replacement of features) that would violate these constraints since its execution would result in an inconsistent constraint store. The semantics of QFLAN (see Section 3.8) ensures that all executions will end up with a consistent configuration if the process (the procedural part, defined below) begins with a consistent constraint store.

Quantitative constraints must be enumerated under the `quantitative constraints` block. The full specification of the quantitative constraints of our running example is depicted in Listing 7.

```
1  begin quantitative constraints
2    { price(Bike) < 600 }
3    { weight(Bike) < 15 }
4    { load(Bike) < 100 }
5  end quantitative constraints
```

Listing 7. Quantitative constraints of the running example

*Action constraints*

QFLAN admits a class of *action constraints*, reminiscent of featured transition systems (FTS) [6].

Each action $a$ can have associated a constraint $\mathtt{do}(a) \to p$, where $p$ is a Boolean formula whose basic predicates are the same as for quantitative constraints and, in addition, predicates of the form $\mathtt{has}(f)$. In our DSL, the default action constraint for each feature $f$ is $\mathtt{do}(f) \to \mathtt{has}(f)$ and it need not be specified. It ensures that in order to use a feature, it must first be installed.

Action constraints act as a kind of guards to either permit or forbid the execution of actions. In our running example, action constraints are used to forbid selling bikes that cost less than 250 euros (C4) and to forbid dumping bikes that cost more than 400 euros (C5). These constraints can be found in Listing 8, showing that action constraints must be enumerated under the `action constraints` block.

```
1  begin action constraints
2    do(sell) -> { price(Bike) > 250 }
3    do(dump) -> { price(Bike) < 400 }
4  end action constraints
```

Listing 8. Action constraints of the running example

We recall that in an FTS, transitions are labeled with actions and with Boolean constraints that work similarly as action constraints, i.e. as guards that impose conditions on the executability of the specific transition. However, action constraints of QFLAN apply to *all* transitions with the same action whereas a condition in FTS applies to *one* instance of an action (i.e. a transition). Grouping the common executability conditions of one action in a dedicated block instead of scattered all over the behavioral specification provides more compact and declarative specifications that are easier to read and to maintain. Specific conditions of the executability of an action can still be specified via queries, as we shall see.

*Constraint semantics*

We are now ready to formally define constraint consistency and constraint entailment, which are crucial in the semantics of QFLAN models.

A constraint system is defined by a set of constraint variables $\chi = \{\vartheta_1, \ldots, \vartheta_n\}$, their domains $\Delta = \{\Delta_1, \ldots, \Delta_n\}$ and a set of constraints $c_1(\vec{\vartheta_1}) \oplus \cdots \oplus c_m(\vec{\vartheta_m})$, where each constraint $c_i$ imposes a restriction on the values of the variables in the vector of variables $\vec{\vartheta_i}$. Thus, a constraint $c_i(\vartheta_{i_1} \ldots \vartheta_{i_k})$ can just be seen as a subset of $\Delta_{i_1} \times \cdots \times \Delta_{i_k}$, i.e. those combinations of values for the variables that the constraint admits.

In QFLAN, the set of constraint variables $\chi$ is the union of the feature propositions $\mathtt{has}(f)$ for all features $f \in \mathcal{F}$, the value of feature predicates $p(f)$ for all predicates $p \in \mathcal{P}$ and all features $f \in \mathcal{F}$, and all user variables in $\mathcal{V}$. The domains of such variables are the Booleans (for feature propositions) and the real numbers (for the rest). The constraints $c_i$ are the ones specified in $\mathcal{C}$ plus the actual values of variables and feature propositions. All in all, they essentially specify which combinations of features, feature attributes and variable values are admitted. In the rest of the section, we fix a constraint system defined by $\chi$, $\Delta$ and $C$.

The typical problems for a constraint system are: (i) *constraint solving*, i.e. finding an assignment $\theta : \chi \to \Delta$ with $\theta(\vartheta_i) \in \Delta_i$ such that each constraint $c_j(\vec{\vartheta_j})$ is satisfied (denoted by $\theta \models c_j(\vec{\vartheta_j})$ here); (ii) *consistency checking*, i.e. merely determining if such an assignment exists; (iii) *constraint checking* or *constraint evaluation*, i.e. checking the consistency of a given assignment $\theta$; and (iv) *constraint entailment*, i.e. checking if a constraint system admits more assignments than another.

Our approach focuses on consistency checking, constraint checking and constraint entailment. Formally, consistency checking is defined as follows.

**Definition 5 (consistency).** Let $C = c_1(\vec{\vartheta_1}) \oplus \cdots \oplus c_m(\vec{\vartheta_n})$ be a constraint store. We say that $C$ is consistent, written $consistent(C)$, iff there is a substitution $\theta : \chi \to \Delta$ such that

$$\bigwedge_{i=1\ldots n} \theta \models c_i(\vec{\vartheta_i}) \equiv true$$

Checking a constraint $C$ for given assignment is just checking if $\theta \models c(\vec{\vartheta})$ for all constraints $c(\vec{\vartheta}) \in C$. This can be done by simply applying the assignment as a substitution on all constraints, thus obtaining a set of Boolean propositions, and evaluating the conjunction of such propositions. We denote by $\theta(C)$ the thus obtained Boolean proposition.

Constraint entailment is formally defined as follows.

**Definition 6 (entailment).** Let $C = c_1(\vec{\vartheta_1}) \oplus \cdots \oplus c_n(\vec{\vartheta_n})$ and $C' = c'_1(\vec{\vartheta_1}) \oplus \cdots \oplus c'_m(\vec{\vartheta_n})$ be two constraint stores.

We say that $C$ entails $C'$, written $C \vdash C'$, iff for all substitutions $\theta$:

$$\bigwedge_{i=1\ldots n} \theta \models c_i(\vec{\vartheta_i}) \rightarrow \bigwedge_{j=1\ldots m} \theta \models c'_j(\vec{\vartheta_j})$$

A constraint $C$ can encode an assignment on all variables in $\chi$ by using constraints of the form $\vartheta_0 = v_0 \oplus \cdots \oplus \vartheta_n = v_n$. When this is the case we denote such assignment by $\theta_C$. In such cases $consistent(C)$ can be reduced to evaluating $\theta_C(C)$, and $C \vdash C'$ for a consistent constraint $C$ can be reduced to evaluating $\theta_C(C) \wedge \theta_C(C')$.

### 3.7 Behavior

The procedural part of QFLAN is provided by specifying a set of processes. QFLAN offers two ways of specifying processes: (i) a process-algebraic specification language based on the original versions of the QFLAN family, and (ii) a state-machine specification language, inspired by the diagrammatic process descriptions used in this paper. As we shall see, the second process description language can be seen as syntactic sugar. It provides a higher-level specification language more amenable for visual representations.

#### Process-algebraic specifications

The behavior $\mathcal{B}$ in a QFLAN specification can be provided in a `processes` block as a set of process definitions `process` $X = P$, where $X$ is a process name and $P$ is a process expression. Process expressions are formally defined as follows.

**Definition 7 (valid process expressions).** Let $\mathcal{S}$ be a QFLAN specification and let $\mathcal{X}$ be a set of process names. The set of $\mathcal{B}$-valid process expressions is defined by the following grammar:

$$
\begin{array}{llll}
P, Q & ::= & \texttt{nil} & \text{(empty process)} \\
 & | & (a, r, u).X & \text{(action and invocation)} \\
 & | & (a, r, u).P & \text{(action and continuation)} \\
 & | & P + Q & \text{(choice)} \\
 & | & P \parallel Q & \text{(parallel composition)}
\end{array}
$$

where $X \in \mathcal{X}$, $a \in \mathcal{A}$, $r$ is a float-valued rate, and $u$ is a set of memory updates of the form $x = e$, where $x \in \mathcal{V}$ is a variable and $e$ is a float expression (over variables in $\mathcal{V}$).

Basic processes can consist of the empty process `nil`, or a single (rated) action with a memory update $u$ followed by the invocation of a process named $X$ or a process $P$.

Action rates are static and allow the modeler to specify probabilistic aspects of product behavior (e.g. the behavior of the user of a product, failure rates of the components of a product or the likelihood of installing a certain feature at a specific moment). A memory update allows the modeller to specify changes on the value of variables. Processes can be combined by non-deterministic choice or in parallel. Prefixing of process invocation is imposed to avoid infinite branching.

#### State-machine specifications

The behavior $\mathcal{B}$ in a QFLAN specification can also be specified with `processes diagram` blocks, each of which describes a state machine (like the one in Fig. 3) whose basic operation is that of executing actions that may change the state of the machine and the values of variables. Process diagram definitions are of the form `begin process` $X$ $D$ `end process`, where $X$ is a process name and $D$ is its diagram definition.

**Definition 8 (process diagram definition).** Let $\mathcal{S}$ be a QFLAN specification. A process diagram definition $D$ in $\mathcal{B}$ is defined by a pair $\langle S, T \rangle$ where

- $S$ is a finite set of user-defined states;
- $T$ is a finite set of transitions $\langle s, (a, r, u), s' \rangle$, where $s \in S$ is the source state, $s' \in S$ is the target state, $a \in \mathcal{A}$ is an action, $r$ is a float-valued rate, and $u$ is a set of variable updates.

#### From state-machines to processes

State-machine specifications can be seen as syntactic sugar for processes. Indeed, every state-machine specification can be translated to a process-algebraic specification. The following definition formalizes the translation.

**Definition 9 (process of a state-machine).** Let $D = \langle S, T \rangle$ be a process diagram definition. The process-algebraic specification for $D$ is defined as the set of process definitions $\{X_s = P_S \mid s \in S\}$ where

$$P_s = \sum_{\langle s, (a, r, u), s' \rangle \in T} (a, r, u).X_{s'}$$

### 3.8 Operational semantics

The semantics of a specification formalizes how a model behaves and changes over time. The semantics of QFLAN is defined over models, which are tuples $\langle C \triangleright P \rangle$ made of a constraint store $C$ and a process $P$. While we could thus consider the general case for $C$ which would possibly define a partially defined product (i.e. a family of products) with some variables and attributes being unspecified, the statistical simulator works for completely defined single products only. For this reason, we restrict here to fully specified models where each constraint store $C$ uniquely determines an assignment to all of its constraint variables.

**Definition 10 (model).** Let $\mathcal{S}$ be a QFLAN specification. A model of $\mathcal{S}$ is a tuple $\langle C \triangleright P \rangle$ where:

- $C$ is a consistent constraint store composed of:

  1) $\mathcal{C}$, i.e. all constraints of the specification are part of the store;
  2) For all $f \in \mathcal{F}$, either $\texttt{has}(f)$ or $\neg\texttt{has}(f)$, i.e. the constraint store $C$ univocally defines a single product;
  3) For all predicates $p \in \mathcal{P}$ and all features $f \in \mathcal{F}$, the set of constraints $p(f) = v$, where $v$ is determined as explained in Section 3.2;
  4) For all variables $v \in \mathcal{V}$, a constraint $x = v$ for some $v \in \mathbb{R}$.

- $P$ is a $\mathcal{B}$-valid process expression.

$$(\textsc{Act}) \; \frac{executable(C, a) \qquad consistent(u(update(C, a)))}{\langle S \rhd (a, r, u).P \rangle \xrightarrow{r} \langle u(update(C, a)) \rhd P \rangle}$$

$$(\textsc{Or}) \; \frac{\langle C \rhd P \rangle \xrightarrow{r} \langle C' \rhd P' \rangle}{\langle C \rhd P + Q \rangle \xrightarrow{r} \langle C' \rhd P' \rangle} \qquad (\textsc{Par}) \; \frac{\langle C \rhd P \rangle \xrightarrow{r} \langle C' \rhd P' \rangle}{\langle C \rhd P \parallel Q \rangle \xrightarrow{r} \langle C' \rhd P' \parallel Q \rangle}$$

$$executable(C, a) = \begin{cases} false & \text{if } C = C' \oplus (\mathtt{do}(a) \to C'') \text{ and } C' \not\vdash C'' \\ false & \text{if } a = \mathtt{install}(f) \text{ and } C = C' \oplus \mathtt{has}(f) \\ false & \text{if } a = \mathtt{uninstall}(f) \text{ and } C = C' \oplus \neg\mathtt{has}(f) \\ false & \text{if } a = \mathtt{replace}(f, g) \text{ and } (C = C' \oplus \neg\mathtt{has}(f) \text{ or } C = C'' \oplus \mathtt{has}(g)) \\ false & \text{if } a = \mathtt{ask}(C') \text{ and } C \not\vdash C' \\ true & \text{otherwise} \end{cases}$$

$$update(C, a) = \begin{cases} C' \oplus \mathtt{has}(f) & \text{if } a = \mathtt{install}(f) \text{ and } C = C' \oplus \neg\mathtt{has}(f) \\ C' \oplus \neg\mathtt{has}(f) & \text{if } a = \mathtt{uninstall}(f) \text{ and } C = C' \oplus \mathtt{has}(f) \\ C' \oplus \neg\mathtt{has}(f) \oplus \mathtt{has}(g) & \text{if } a = \mathtt{replace}(f, g) \text{ and } C = C' \oplus \mathtt{has}(f) \oplus \neg\mathtt{has}(g) \\ C & \text{otherwise} \end{cases}$$

Fig. 2. Operational semantics

Let $\mathcal{M}$ denote the set of all models for a specification $\mathcal{S}$. The *initial* model is the model $\langle C \rhd P \rangle$ such that the initial product defined by $C$ and the process $P$ are defined in the `init` block of the specification, and the values of variables specified in $C$ are taken from the specification of the `variables` block.

The fact that the constraint store $C$ univocally defines a single product (cf. Definition 10(2)) implies that in the presence of a cross-tree constraint $f$ requires $g$, whenever both features are present in a product (configuration), then it must always be the case that feature $g$ was installed first.

In our running example, we assume that bikes are pre-configured, containing precisely one of the alternative sub-features from each of the mandatory features `Wheels` and `Frame`. For example, an initial product from the bikes product line can contain the feature set {`AllYear`, `Diamond`}.

*Transition system semantics*

The operational semantics of models is formalized in terms of the state transition relation $\to \; \subseteq \mathbb{N}^{\mathcal{M} \times \mathbb{R}^+ \times \mathcal{M}}$ defined in Fig. 2. Note that we use multisets of transitions to deal with the possibility of multiple instances of a transition. Technically, such a reduction relation is defined in structural operational semantics, i.e. by induction on the structure of models, up to process invocation and to associativity, commutativity and identity of choice and parallel composition, formalized by the structural congruence $\equiv \; \subseteq \mathcal{M} \times \mathcal{M}$ defined by

$$P + (Q + R) \equiv (P + Q) + R \qquad P + Q \equiv Q + P$$
$$P \parallel (Q \parallel R) \equiv (P \parallel Q) \parallel R \qquad P \parallel Q \equiv Q \parallel P$$
$$P + \emptyset \equiv P \qquad P \parallel \emptyset \equiv P$$
$$P \equiv P[^Q/_X] \text{ if } X = Q \in \mathcal{B}$$

As usual, the transition rules in Fig. 2 are expressed as a set of premises (above the line) and a conclusion (below the line). The transition relation implicitly defines a labeled transition system (LTS) $\langle \mathcal{M}, \to \rangle$, whose states are models and whose transitions are labeled with rates. Given a state $s \in \mathcal{M}$, we denote the total outgoing rate from $s$ as:

$$\mathbf{out}(s) = \sum_{(s, r, s') \in \to} r$$

The rule $\textsc{Act}$ allows a process to execute an action if:

- the executability constraints for $a$ in $C$ allow so (formalized by $executable(C, a)$); and
- if the resulting constraint store after its modification $u(update(C, a))$ by the action $a$ and the update $u$ is consistent.

Action constraints can impose executability conditions on all actions. A typical action constraint is $\mathtt{do}(a) \to \mathtt{has}(f)$, i.e. action $a$ is subject to the presence of feature $f$. The installation, removal and replacement of features have additional constraints. An action $\mathtt{ask}(C')$ works as in concurrent constraint programming [18], i.e. it has an additional executability constraint (the entailment of $C'$) that may block the action.

Note that all actions may have a set of updates $u$ of variables. The effect of applying such an effect on $C$ (denoted $u(C)$) is the expected one, i.e. an update $x = e$ evaluates $e$ in $\theta_C$ and replaces $x = v$ by $x = \theta_C(e)$ in $C$. In addition, the store actions $\mathtt{install}(f)$, $\mathtt{uninstall}(f)$ and $\mathtt{replace}(f, g)$ update the constraint store as defined in Fig. 2, i.e. by adding, removing or replacing features.

Rules $\textsc{Or}$ and $\textsc{Par}$ are standard, formalizing non-deterministic choice and interleaving parallel composition, respectively. Note that non-determinism introduced by choice and parallel composition is probabilistically resolved in the probabilistic semantics (described later).

We note three ways to include a feature $f$ in a product configuration. First, an *explicit*, *declarative* way is to include the proposition $\mathtt{has}(f)$ in the initial store; this is the way to include core features (i.e. globally mandatory features). Second, an *implicit*, *declarative* given by the closure mechanism defined in Section 3.2. Third, a *procedural* way is to dynamically install $f$ at runtime, possibly by replacement.

*DTMC semantics*

We recall the definition of a discrete-time Markov chain (DTMC).

**Definition 11 (DTMC).** A DTMC is a tuple $\langle \Gamma, \Pi \rangle$ where:

- $\Gamma$ is a set of states;
- $\Pi : \Gamma \to [0, 1]$ is a probability transition function, i.e. such that for all $s \in \Gamma$, $\sum_{s' \in \Gamma} \Pi(s, s') = 1$.

It is straightforward to obtain a discrete-time Markov chain (DTMC) from an LTS as above by normalising the rates

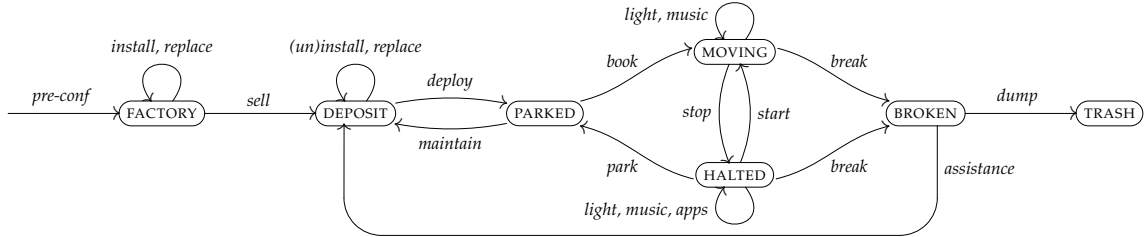

Fig. 3. Sketch of bike-sharing behavior

into [0..1] such that in each state, the sum of the rates of its outgoing transitions equals one.

Formally, this is defined as follows

**Definition 12 (DTMC of rate-based LTS).** Let $\mathcal{T} = \langle \mathcal{M}, \rightarrow \rangle$ be a rate-labeled transition system. The DTMC of $\mathcal{T}$ is $\langle \mathcal{M}, \Pi \rangle$ where, for each pair of states $s, s' \in \mathcal{M}$, the probability transition function $\Pi$ is defined by

$$\Pi(s, s') = \begin{cases} \frac{\sum_{(s, r, s') \in \rightarrow} r}{\mathbf{out}(s)} & \text{if } \mathbf{out}(s) > 0 \\ 1 & \text{if } \mathbf{out}(s) = 0 \text{ and } s = s' \\ 0 & \text{otherwise} \end{cases}$$

In the resulting DTMC, the rates of the transition system are normalized into probabilities: the probability that the transition is taken from its source state. States without outgoing transitions are enriched with self-loops with rate equal to 1.

Recall that we advocate the use of SMC since it uses on-the-fly generated simulations of the DTMC, which in general is too large to be generated explicitly.

## 4 RUNNING EXAMPLE REVISITED: DYNAMICS

We now illustrate the presentation of the process specification with our running example. The behavior associated to our bikes product line is based on a bike-sharing scenario that we abstracted from the bike-sharing system *CicloPi* (cf. Section 2) with some additional behavior concerning not yet realized features, such as the use of electric bikes and the possible runtime installation of apps. These are features that have been envisioned for 4th generation bike-sharing systems [38], [39], some of which (e.g. Tablet, Engine and Battery) have become reality in the recently deployed *Bycyklen* bike-sharing system in Copenhagen.

The dynamics of our running example is given in terms of one process only, sketched in Fig. 3. For simplicity, the action rates are not depicted in Fig. 3, but they can be found in Listing 9 which provides the entire specification of the process. In FACTORY (e.g. of Bicincittà), features may be installed or replaced (e.g. different wheels or a different frame). At a certain point, the configured bike may be sold (as part of a bike-sharing system), but only if it costs at least 250 euro (to satisfy constraint (C4) on action *sell*), after which it arrives in the DEPOSIT (e.g. of PisaMo). It may then be ready to be deployed as part of the bike-sharing system run from this deposit, or it may first need to be further fine-tuned by (un)installing or replacing factory-installed features. Once it is deployed, it results PARKED in one of the docking stations of the bike-sharing system (e.g. *CicloPi*).

A user may book a PARKED bike, resulting in a MOVING bike. While biking, a user may decide to listen to music or switch on the light, in case the corresponding features have been installed. If a user wants to consult one of the apps (a map, a navigator or a guide), then (s)he first needs to stop biking, resulting in a HALTED bike, from where (s)he may consult an app, before eventually start to bike again or park the bike in a docking station. Unfortunately, the bike may also break, resulting in a BROKEN bike. Hence, assistance from the bike-sharing system exploiter arrives. If the bike can be fixed, it is brought to the DEPOSIT. If the damage is too severe, and the bike has a price of at most 400 euros (to satisfy constraint (C5) on action *dump*), then we dump the bike in the TRASH. At regular intervals, assistance from the bike-sharing system exploiter takes a PARKED bike to the DEPOSIT for maintenance.

As said before, the detailed process specification of the case study can be found in Listing 9. Note the tight correspondence between Listing 9 and its graphical representation in Fig. 3. In particular, it contains one state per node in Fig. 3, and a set of transitions per edge in Fig. 3. When the system is in state `factory` we face a choice, weighted by the rates, among three main activities:

1. Sell the bike and send it to the `deposit` (with rate 8). This action can only be executed if (C4) is respected;
2. Install optional features and iterate on `factory`. The installations are performed only if the constraints are not violated;
3. Replace pre-installed child features of the mandatory (abstract) features `Wheels` or `Frame`. Again, respecting the constraints.

Note that in (2) we assume that `Music` is the feature installed with higher probability, followed by `MapsApp`, `Dynamo` and `Light`. Note that the semantics of QFLAN forbids the re-installation of installed features (i.e. a product is a set of features, and not a multiset). In (3), we favor the replacement of `Winter` or `Summer` wheels by `AllYear` ones. A frame may be changed as well, but with lower probability.

State `deposit` is similar to `factory`. Clearly, `deposit` differs by the possibility to perform an action `deploy` leading to process `parked`. In addition, `deposit` may also uninstall features, so as to allow for customization. Optional features can be installed and uninstalled with the same rate by `deposit`, except for `Engine`, `Battery` and `Dynamo`, which are uninstalled with a lower rate to penalize their occurrence. This modeling choice is justified by the fact that it is reasonable to assume that uninstalling such features may cost more than installing them. In addition, we assume that the frame identifies the bike that was sold, and thus it cannot be modified in `deposit`.

```
1   begin processes diagram
2     begin process bikesProcess
3       states = factory , deposit , parked , moving ,
4                halted , broken , trash
5       transitions =
6       // Sell bike from factory
7       factory -(sell , 8)-> deposit ,
8       // Install optional features of bike in factory
9       factory -(install(GPS) , 6)-> factory ,
10      factory -(install(MapsApp) , 10)-> factory ,
11      factory -(install(NaviApp) , 6)-> factory ,
12      factory -(install(GuideApp) , 3)-> factory ,
13      factory -(install(Music) , 20)-> factory ,
14      factory -(install(Engine) , 4)-> factory ,
15      factory -(install(Battery) , 4)-> factory ,
16      factory -(install(Dynamo) , 10)-> factory ,
17      factory -(install(Light) , 10)-> factory ,
18      factory -(install(Basket) , 8)-> factory ,
19      // Replace child features of mandatory features of
20      // bike in factory
21      factory -(replace(AllYear , Summer) , 5)-> factory ,
22      factory -(replace(AllYear , Winter) , 5)-> factory ,
23      factory -(replace(Summer , AllYear) , 10)-> factory ,
24      factory -(replace(Summer , Winter) , 5)-> factory ,
25      factory -(replace(Winter , Summer) , 5)-> factory ,
26      factory -(replace(Winter , AllYear) , 10)-> factory ,
27      factory -(replace(Diamond , StepThru) , 3)-> factory ,
28      factory -(replace(StepThru , Diamond) , 3)-> factory ,
29      // Deploy bike from deposit
30      deposit -(deploy , 10 , { deploys = (deploys + 1) })->
                parked ,
31      // Install optional features of bike in deposit
32      deposit -(install(GPS) , 6)-> deposit ,
33      deposit -(install(MapsApp) , 10)-> deposit ,
34      deposit -(install(NaviApp) , 6)-> deposit ,
35      deposit -(install(GuideApp) , 3)-> deposit ,
36      deposit -(install(Music) , 20)-> deposit ,
37      deposit -(install(Engine) , 4)-> deposit ,
38      deposit -(install(Battery) , 4)-> deposit ,
39      deposit -(install(Dynamo) , 10)-> deposit ,
40      deposit -(install(Light) , 10)-> deposit ,
41      deposit -(install(Basket) , 8)-> deposit ,
42      // Uninstall optional features of bike in deposit
43      deposit -(uninstall(GPS) , 6)-> deposit ,
44      deposit -(uninstall(MapsApp) , 10)-> deposit ,
45      deposit -(uninstall(NaviApp) , 6)-> deposit ,
46      deposit -(uninstall(GuideApp) , 3)-> deposit ,
47      deposit -(uninstall(Music) , 20)-> deposit ,
48      deposit -(uninstall(Engine) , 1)-> deposit ,
49      deposit -(uninstall(Battery) , 2)-> deposit ,
50      deposit -(uninstall(Dynamo) , 3)-> deposit ,
51      deposit -(uninstall(Light) , 10)-> deposit ,
52      deposit -(uninstall(Basket) , 8)-> deposit ,
53      // Replace child features of mandatory features
54      // (Frame cannot be changed)
55      deposit -(replace(AllYear , Summer) , 5)-> deposit ,
56      deposit -(replace(AllYear , Winter) , 5)-> deposit ,
57      deposit -(replace(Summer , AllYear) , 10)-> deposit ,
58      deposit -(replace(Summer , Winter) , 5)-> deposit ,
59      deposit -(replace(Winter , Summer) , 5)-> deposit ,
60      deposit -(replace(Winter , AllYear) , 10)-> deposit ,
61      // Replace Battery with Dynamo, if battery is not used
62      deposit -(replace(Battery , Dynamo) , 1)-> deposit ,
63      // Behavior of deployed bike
64      parked -(book , 10)-> moving ,
65      parked -(maintain , 1)-> deposit ,
66      moving -(stop , 5)-> halted ,
67      moving -(break , 1)-> broken ,
68      moving -(Music , 20)-> moving ,
69      moving -(Light , 20)-> moving ,
70      halted -(start , 5)-> moving ,
71      halted -(break , 1)-> broken ,
72      halted -(Music , 20)-> halted ,
73      halted -(GPS , 10)-> halted ,
74      halted -(GuideApp , 10)-> halted ,
75      halted -(MapsApp , 10)-> halted ,
76      halted -(NaviApp , 10)-> halted ,
77      halted -(Light , 10)-> halted ,
78      broken -(assistance , 10)-> deposit ,
79      broken -(dump , 1 , { trashed = 1 })-> trash
80    end process
81  end processes diagram
```

Listing 9. The process defining the dynamics of the running example

The final action that `deposit` can perform is an interesting one: feature `Battery` can be replaced with the much cheaper `Dynamo`. According to the semantics of QFLAN, this action is performed only if no subfeatures of `CompUnit` or of the `Engine` are currently installed (cf. Fig. 1). This is useful to reduce costs and weight, in case some previously installed feature requiring the battery has by now been uninstalled.

The remaining states `parked`, `moving`, `halted`, `broken` and `trash` are rather simple and are faithful to their informal description above. It is worth to discuss the transitions in Line 30 and Line 79 of Listing 9. In both cases, the transition also updates a *variable*, `deploys` and `trashed`, respectively, used to record the number of times that the bike has been deployed or if it is trashed.

Note that `factory` is a pure (pre-)configuration state, while `deposit` is not. In fact, parked bikes can be brought back into the deposit, and thus features can be (un)installed or replaced at runtime. This is an example of a staged configuration process, in which some optional features are bound at runtime rather than at (pre-)configuration time.

The initial system configuration is specified in the `init` block. This is provided in Listing 10 for our running example, showing that the initially installed features are `Diamond` and `AllYear`, while the dynamics are given by the process `bikesProcess` (starting in state `factory`, the first state defined in the corresponding `process` block in Lines 3–4 of Listing 9). In case the dynamics of the model under analysis are given in terms of more than one process, this can be specified in `initialProcesses` of the `init` block by listing all required processes (separated by the character `|`). The state of each process will be maintained, and one transition among all those outgoing from all such states will be probabilistically chosen at each step.

```
1  begin init
2    installedFeatures = { Diamond , AllYear }
3    initialProcesses = bikesProcess
4  end init
```

Listing 10. Initial system configuration of the running example

In particular, the modeller is required to start from an initial configuration that satisfies all constraints. This is enforced by an automatic static analysis offered by our tool framework that checks for the validity of all constraints in the initial configuration, and lists those which failed.

The full QFLAN specification of the case study can be found at http://github.com/qflanTeam/QFLan/wiki

## 5 THE QFLAN TOOL

This section presents the QFLAN tool, a multi-platform application based on Eclipse which enables the modeling and analysis of QFLAN specifications. The tool is available together with installation and usage instructions at http://github.com/qflanTeam/QFLan/wiki.

Figure 4 depicts the architecture of the QFLAN tool framework. It is organized in the GUI layer, devoted to modeling aspects, and the core layer, offering support for the analysis of QFLAN specifications.

The components of the GUI layer are shown in Fig. 5. The most notable one is a text editor with editing support typical of modern integrated development environments (auto-completion, syntax and error highlighting, etc.)

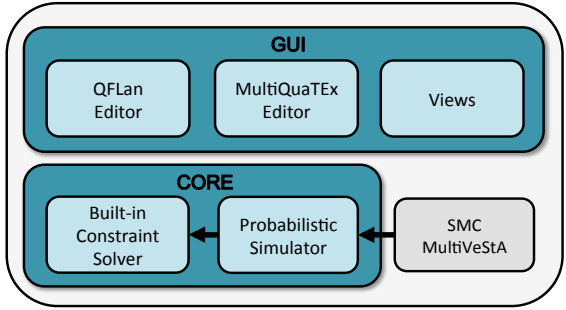

Fig. 4. The architecture of the QFLAN tool

developed within the XTEXT framework (top-middle of Fig. 5). The XTEXT grammar of QFLAN and the rest of the source code are available at http://github.com/qflanTeam/QFLan/wiki. For instance, the editor is able to promptly highlight incorrect feature diagrams (e.g. including features with more than one parent or abstract features without descendants) or incorrect feature predicates (e.g. multiple values assigned to one feature). The editor also offers support for the MultiQuaTEx query language, used to analyse QFLAN specifications (cf. Section 6). In addition, the GUI layer offers a number of views, including: a *console view* to display diagnostic information (bottom of Fig. 5); a *project explorer* to handle different QFLAN specifications (left of Fig. 5); and a *plot view* to display analysis results (top-right of Fig. 5).

The main component of the core layer is the probabilistic simulator of QFLAN models. Intuitively, we obtain probabilistic simulations of a QFLAN model by executing it step-by-step starting from the initial configuration specified by the modeler. At each iteration we compute the set of admissible transitions, and select one of them according to the probability distribution resulting from normalizing the rates of the generated transitions. In particular, checking if an action is admissible amounts to checking whether it violates any constraint.

The simulator implements a number of optimizations in order to improve performance by reusing computations performed in previous steps whenever possible. For example, we re-check the admissibility of an action only if the constraint store has been modified in a way that could affect its admissibility.

Contrary to previous prototypes, our new tool does not use the SMT solver Z3. The main reason is that the analyses we consider requires us to work with fully specified models, i.e. models $\langle C \rhd P \rangle$ such that the assignment $\theta_C$ is well defined. This is because the observations on a model must be deterministic. Hence, as explained in Section 3.6, the constraint problems that the QFLAN interpreter needs to solve can be reduced to a constraint checking problem, i.e. to the evaluation of Boolean propositions.

## 6 STATISTICAL ANALYSIS OF QFLAN MODELS

In order to perform automated quantitative analysis of QFLAN specifications, we integrated the distributed statistical model checker MULTIVESTA [28] within the QFLAN tool framework.

### 6.1 MultiVeStA

MULTIVESTA can easily be integrated with any formalism that allows probabilistic simulations and it has already been used to analyze a wide variety of systems, including contract-oriented middlewares [40], opportunistic network protocols [41], online planning [42], crowd-steering [43], public transportation systems [44], [45], volunteer clouds [46] and swarm robotics [47].

Within the QFLAN tool, MULTIVESTA can be used to obtain statistical estimations of quantitative properties of QFLAN specifications. MULTIVESTA provides such estimations by means of distributed analysis techniques known from statistical model checking (SMC) [48], [49].

Classical SMC allows one to perform analyses like "is the probability that a property holds greater than a given threshold?" or "what is the probability that a property is satisfied?". In addition, MULTIVESTA also allows one to estimate the expected values of properties that can take on any value from $\mathbb{R}$, like "what is the average cost/weight/load of products configured according to an SPL specification?". Estimations are computed as the mean of $n$ samples obtained from $n$ independent simulations, with $n$ large enough (but minimal) to grant that the size of the $(1 - \alpha) \times 100\%$ *confidence interval* (CI) is bounded by $\delta$. In other words, if MULTIVESTA estimates the value of a property as $\overline{x} \in \mathbb{R}$, then with probability $(1 - \alpha)$ its actual expected value belongs to the interval $[\overline{x} - \delta/2, \overline{x} + \delta/2]$. A CI is thus specified in terms of two parameters: $\alpha$ and $\delta$.

### 6.2 Property specification

MULTIVESTA allows to estimate properties like the average of real-valued observations on the model behavior. As depicted in Fig. 4, the QFLAN tool offers an editor for property specifications. The novel property specification language provides a high-level abstraction of MULTIQUATEX [28].

As done in [16], the analysis capabilities of our framework are exemplified using three families of properties of interest to our case study:

$(P_1)$  Average price, weight and load of a bike when it is deployed for the first time, or as time progresses;

$(P_2)$  For each of the 15 concrete features that appear as leaves in the feature model of Fig. 1, the probability to have it installed when a bike is deployed for the first time, or as time progresses;

$(P_3)$  The probability for a bike to be dumped.

When analyzed at the first deployment of a bike, $P_1$ and $P_2$ are useful for studying a sort of *initial scenario*, in order to estimate the required initial investments and infrastructures. For instance, bikes with a high price and a high load (i.e. with a high technological footprint) or equipped with a battery might require docking stations with specific characteristics, or they might have to be collected for the night to be stored safely. Instead, analyzing $P_1$ and $P_2$ as time progresses provides an indication of how those values evolve, e.g. to estimate the average value in euros of a deployed bike and the monetary consequences of its loss.

From a more general perspective, properties like $P_2$ and $P_3$ measure how often (on average) a feature is actually installed in a product from a product line or how often (on average) a bike is dumped in the trash. The outcome of the former provides important information for those responsible for the production or programming of a specific feature or system module.

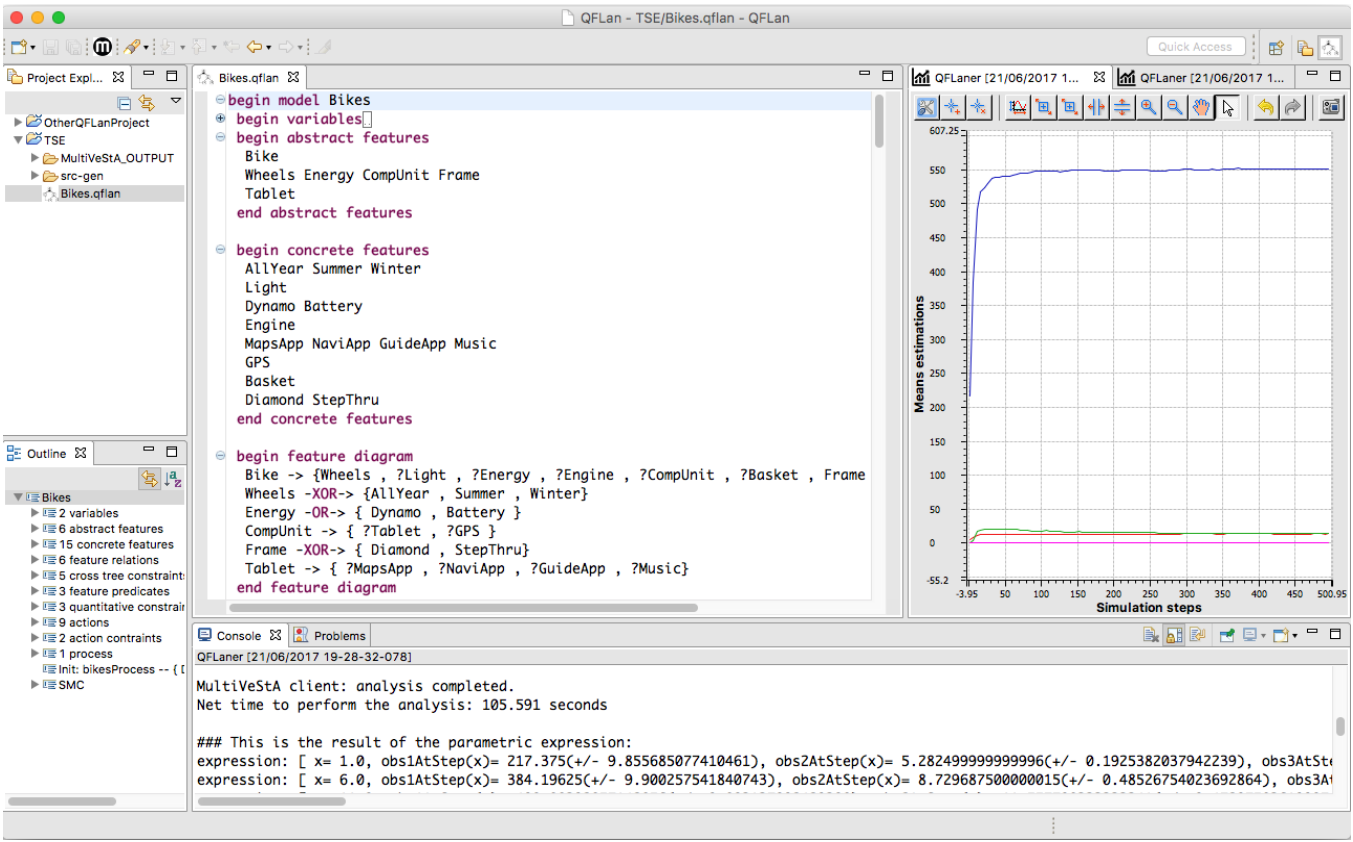

Fig. 5. A screenshot of the QFLan tool

*Specification of $P_1$ and $P_2$ at first deployment*

Listing 11 depicts the code snippet to be specified within the QFLan tool in order to evaluate $P_1$ and $P_2$ at a bike's first deployment. These are examples of state properties to be evaluated in a given state of each performed simulation, i.e. the first state met that satisfies the condition specified by the keyword `when`, shown in Line 2. As discussed in Section 4, `deploys` is a variable that is increased when the bike is deployed from the deposit to a parking lot. The `when` clause can be given as a Boolean expression involving the presence/absence of a given feature, and (in)equations on the value of a predicate or of a variable (e.g. `deploys > 0` in Line 2). Property $P_1$ is specified in Lines 3–4, where we specify that we want to study the average price, weight and load of bikes. In addition, in Line 4, we also query the expected number of simulation `steps` to perform the first deployment. In square brackets we specify the $\delta$ to be used for these properties. Instead, Lines 5–7 correspond to $P_2$. In fact, by providing the name of a feature we query the QFLan tool to study the probability of having it installed in the state of interest. In particular, a state property can be any arithmetic expression involving `step`, a feature, a variable or a predicate.

```
begin analysis                                          1
  query = eval when { deploys > 0 } :                   2
    { price(Bike)[delta = 20] , weight(Bike)[delta = 1] ,   3
      load(Bike)[delta = 5] , steps[delta = 1] ,        4
      AllYear , Summer , Winter , GPS , MapsApp ,        5
      NaviApp , GuideApp , Music , Diamond , StepThru ,  6
      Battery , Dynamo , Engine , Basket , Light }       7
  default delta = 0.1                                    8
  alpha = 0.1                                            9
  parallelism = 4                                        10
end analysis                                             11
```

Listing 11. $\mathbf{P}_1$ and $\mathbf{P}_2$ at first deployment

For the state observations of Lines 5–7, the default value of $\delta$ provided in Line 8 is used, while all properties share the same alpha specified in Line 9. Finally, the keyword `parallelism` in Line 10 locally distributes the simulations across four distinct Java processes (which will be allocated on different cores, if possible, by the JVM). All experiments described in this paper use value 4 for the parameter `parallelism`. By design choice, viz. to have a stand-alone easy-to-use application, we do not use the ability of MULTIVESTA to distribute simulations across different machines. However, no technical reason prevents us from extending our tool in this way in the future.

Notably, Listing 11 shows how QFLAN allows one to express more properties at once (in this case 19) which are estimated by MULTIVESTA reusing the same simulations. Furthermore, more queries can be expressed, each with its `when` clause, and list of state observations, again all evaluated reusing the same simulations. We remark that a procedure taking into account that each property might require a different number of simulations is adopted to satisfy the given confidence interval CI.

TABLE 2
Property $P_1$ evaluated at a Bike's first deployment, and average number of steps required for deployment

| Constraints | | Steps to deploy | Feature attributes ($P_1$) | | |
|---|---|---|---|---|---|
| C1 | C2 | | Price | Weight | Load |
| 600 | 15 | 12.19 | 380.92 | 8.10 | 20.92 |
| 800 | 20 | 13.33 | 508.89 | 12.46 | 21.74 |

In Section 7.1 we will see a variant of properties as in Listing 11 with the keyword `when` replaced with `until`. Intuitively, `until` checks that a Boolean property holds in all simulation states met until a given condition holds.

### Encoding of $P_1$–$P_3$ as time progresses

We now discuss how to express variants of $P_1$ and $P_2$ as well as $P_3$ measured as time progresses, demonstrating how to analyze properties upon the variation of a parameter, in this case the number of performed simulation steps. Listing 12 shows the code snippet necessary to analyze such properties. Essentially, the only required change (cf. Line 2 of Listing 11 and Line 2 of Listing 12) is to substitute the keyword `when` with: (i) the parameter of interest, specified by the keyword `for`, followed by a variable, a predicate or a feature (in this case the variable step); and (ii) the values of interest for the parameter (starting `from` an initial value, up `to` a final value, `by` a given increment). This is shown in Line 2, specifying values 1, 6, 11, ..., 496. As we will see, this interval is reasonable, since all studied properties tend to stabilize within this interval. As for Listing 11, Lines 3–5 correspond to properties $P_1$ and $P_2$, like before. Instead, Line 6, corresponding to $P_3$, concerns the probability to dump the bike. In fact, as discussed in Section 4, `trashed` is a variable set to 1 when the bike is dumped.

```
1  begin analysis
2    query = eval for step from 1 to 500 by 5 :
3      { price(Bike)[delta = 20] , weight(Bike)[delta = 1] ,
4        load(Bike)[delta = 5] ,
5        AllYear , Summer , Winter , ... , Basket , Light ,
6        trashed }
7    default delta = 0.1
8    alpha = 0.1
9    parallelism = 4
10 end analysis
```

Listing 12. $P_1$–$P_3$ for varying simulation steps

### 6.3 Statistical analysis of our bikes case study

We now report on the evaluation of the discussed properties. All experiments were performed on a laptop equipped with a 2.4 GHz Intel Core i5 processor and 4 GB of RAM, distributing the simulations among its 4 cores (i.e. setting `parallelism` to 4).

### Evaluation of $P_1$ and $P_2$ at first deployment

The analysis of Listing 11 required 480 simulations, performed in about 8 seconds. In particular, `steps` is the property that required more simulations, viz. 480, while `price` required only 160 simulations. The results are shown in the first row of Tables 2 and 3. Notably, the probability of installing an engine is very low, estimated at 0. Note that, given that the specified confidence interval is $\alpha = 0.1$ and $\delta = 0.1$, the estimated value 0 needs to be interpreted as being in the interval $[0, 0.05]$ with probability 0.9. We guess that this is due to constraints (C1) and (C2), imposing bikes to cost less than 600 euros, and weighing less than 15 kilos. In fact, the estimated average price and weight of bikes at first deployment is 380.92 euros and 8.1 kilos, respectively, while an engine costs 300 euros and weighs 10 kilos. In order to confirm this hypothesis, we analyzed the same property in a new model in which (C1) and (C2) allow bikes to cost at most 800 euros and weigh at most 20 kilos. The results, shown in the second row of Tables 2 and 3, confirm our hypothesis. This analysis thus revealed that the constraints were in disagreement with the quantitative attributes of the features. The latter analysis required 1,200 simulations, performed in about 10 seconds. In this case the estimation of the average price required 1,200 simulations rather than 160 as in the first case. This is because the looser constraints of the latter analysis induce a higher variability of bike prices. In fact, the installation of an engine, the most expensive among the considered features, results in a steep increase of bike prices.

### Evaluation of $P_1$–$P_3$ as time progresses

We evaluated the property of Listing 12 for our case study. We report the results obtained for the model in which (C1) and (C2) bound the price and weight of a bike to 800 and 20, respectively. All such analyses ($19 \times 25$ different properties) were evaluated using the same simulations. Overall, 1,200 simulations were needed, performed in about 80 seconds. The results are presented in four plots in Figure 6: one for the average price (a), one for the average weights and loads (b), one for the probabilities of installing features (c) and one for the probability of dumping the bike (d).

Figure 6a shows that the average price (on the y-axis) of the intermediate bikes generated from the product line at step 1 is 230, hence higher than the 200 euros of the initial configuration (with AllYear and Diamond installed). In particular, it is possible to see an initial fast growth of the price until reaching an average price of about 510 euros, after which the growth slows down, reaching about 540 euros at step 100 and 543 at step 500. This is consistent with our QFLan specification, which has a pre-configuration phase (FACTORY) during which a number of features can be installed, followed by a customization phase (DEPOSIT), where features can be (un)installed and replaced. We recall that FACTORY does not perform any uninstalling, while we note that the uninstalling actions of DEPOSIT do not introduce decrements of the price, on average. A manual inspection of the data revealed that the phase of fast growth slows down between the observed steps 11 and 16. This is consistent with the analysis described in the second row of Table 2, where the average number of steps to complete the first DEPOSIT phase is estimated as being close to 13. In addition, the average price reported in Table 2 at that time step is within the prices observed at steps 11 and 16, respectively 492 and 515. Note, finally, that the probability of a bike to return to the DEPOSIT after its first deployment is quite low. In fact, as specified in Listing 9, PARKED has a transition with rate 10 towards MOVING and one with rate 1 towards DEPOSIT. Thus, on average, the price of bikes is

TABLE 3
Property $P_2$ evaluated at a Bike's first deployment

| Constraints | | Concrete features ($P_2$) | | | | | | | | | | | | | |
|---|---|---|---|---|---|---|---|---|---|---|---|---|---|---|---|
| C1 | C2 | AllYear | Summer | Winter | Light | Dynamo | Battery | Engine | MapsApp | NaviApp | GuideApp | Music | GPS | Basket | Diamond | StepThru |
| 600 | 15 | 0.55 | 0.26 | 0.22 | 0.54 | 0.77 | 0.91 | 0.0 | 0.27 | 0.04 | 0.29 | 0.47 | 0.04 | 0.67 | 0.66 | 0.34 |
| 800 | 20 | 0.57 | 0.24 | 0.20 | 0.59 | 0.74 | 0.89 | 0.44 | 0.24 | 0.06 | 0.27 | 0.50 | 0.10 | 0.62 | 0.73 | 0.27 |

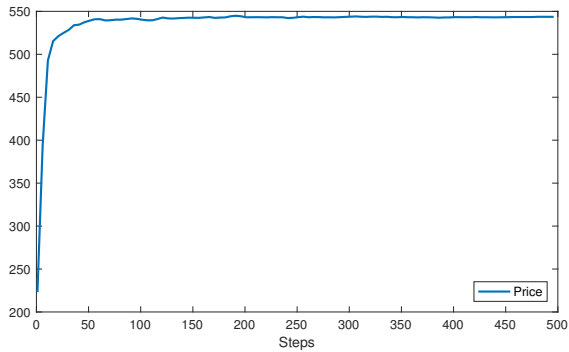

(a) $P_1$ and $P_2$ (price)

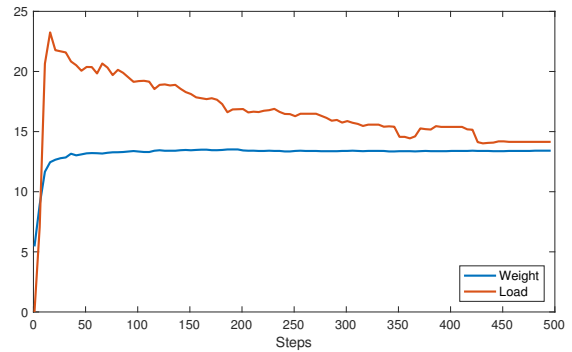

(b) $P_1$ and $P_2$ (weight and load)

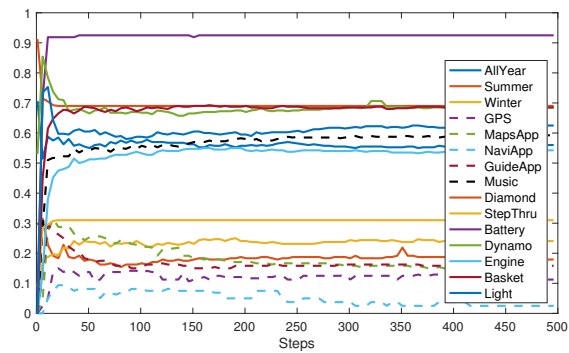

(c) $P_1$ and $P_2$ (installation likelihood)

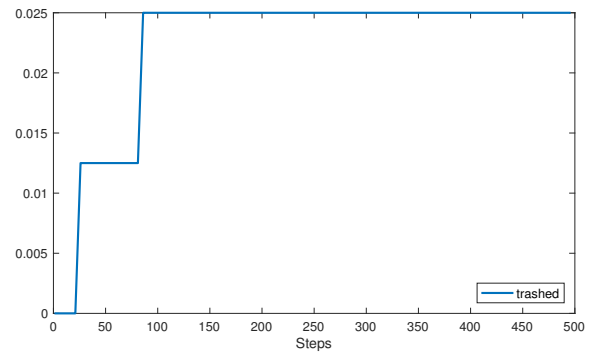

(d) $P_3$ (dumping a bike)

Fig. 6. Results of measuring $P_1$–$P_3$ for the model with constraints C1 = 800 (price) and C2 = 20 (weight)

only slightly affected by (un)installations and replacements performed by successive DEPOSIT phases.

Next to the low probability of actually uninstalling features, there are three more reasons for which uninstalling actions of DEPOSIT do not introduce decrements of the price, on average. First, the features Engine and Battery, which are by far the most expensive ones, are uninstalled with the two lowest rates (1 and 2, respectively). This is because DEPOSIT is about customizations and it is thus unlikely that an engine is uninstalled in the DEPOSIT phase after it has been installed in the FACTORY phase. Second, the features Engine and Battery are involved in two requires cross-tree constraints, which limits the possibility to actually uninstall them (e.g. if CompUnit is installed, then Battery cannot be uninstalled). Third, the features Wheels and Frame, which after Engine and Battery are the most expensive ones, can never be uninstalled but only replaced, since any bike must have wheels and a frame.

Figure 6b shows that weight evolves similarly to price: a first phase of fast growth is followed by a slower growth. As regards the load, instead, after the phase of fast growth up to 23%, it slowly decreases to stabilize around 14%.

It is interesting to observe that load varies for a rather long period after an initial phase of fast growth, while both weight and price (cf. Fig. 6a) quickly stabilize after a short phase of fast growth. This is because load is an attribute of only five features, all of which are among the cheapest features (ranging from 10 to 20 euros) and among those that are uninstalled with the highest rates. In particular MapsApp and Music are uninstalled with rates 10 and 20, respectively, and cost only 10 euros.

As confirmed by Fig. 6c, the probabilities (on the y-axis) for each of the features to be installed evolve similarly to the average price and weight of the generated products, although, clearly, with different scales. It is interesting to note that the pre-installed features AllYear and Diamond have high probability of being installed at step 1, after which the probability decreases during the first 16 steps. The dashed lines refer to all concrete features descending from CompUnit, which are the only ones with non-zero computational load. The slow decrease of load shown in Fig. 6b is due to the slow decrease in having installed MapsApp, NaviApp and GuideApp, which are the features with highest computational load. Instead, Music, whose probability of being installed remains above 0.5 has only 5% of computational load.

Figure 6d shows that bikes are dumped with very low probability. The reason is twofold. First, the transition from BROKEN to TRASH has a much lower rate than the one to DEPOSIT, and similarly for those from MOVING and HALTED to BROKEN (cf. Listing 9). Second, the average price of bikes quickly rises above 400 euros (Fig. 6a) and constraint (C5) prohibits dumping bikes costing more than 400 euros.

# 7 EVALUATION

We have used our tool-supported methodology to model and analyze a number of small case studies in our current and previous work, including classical ones from the SPLE literature such as the coffee vending machine [10], [50], [51], [52], [53], [54], [55] as well as novel ones such as the running example of bikes used here. We report in this section on two additional case studies that witness two particular features of our approach, namely scalability of the analysis and flexibility of the modelling language and its analysis support. Scalability is addressed in Sections 7.1–7.3. First, in Section 7.1 we use the classical example of a product line of elevators to evaluate the scalability of our tool support with respect existing approaches. This case study has been shown to be very demanding in terms of scalability when large sizes of elevator systems are considered (cf., e.g. [34], [56]) and we will demonstrate that we can handle significantly larger instances with respect to existing approaches. Then, in sections 7.2 and 7.3, respectively, we evaluate the impact on the analysis effort of two key characteristics of QFLAN models, namely dynamicity of models in terms of reconfiguration frequency and product variability as imposed by feature constraints. To show flexibility, we model and analyze in Section 7.4 a novel case study that extends a classical example of risk analysis of a safe lock system, thus illustrating the applicability of our approach also in a non-SPL setting.

All the case studies used in this section as well as our bikes running example are available at http://github.com/qflanTeam/QFLan/wiki/Models-from-TSE-submission.

## 7.1 Elevator

The case study we consider here is adapted from the various incarnations of the Elevator product line, originally introduced in [57], which has become a benchmark for SPL analysis (cf., e.g., [12], [23], [34], [56], [58], [59], [60], [61], [62]). This case study is particularly challenging, not so much due to the number of independent, unconstrained features (9, yielding 512 products) but rather due to the need to consider instances with a large number of floors.

The Elevator SPL consists of a number of platform and cabin buttons, one for each platform, which call the elevator. A button that is pressed (chosen non-deterministically) remains pressed until the elevator has served the floor and opened its doors. Serving a floor means opening and closing its doors. We consider the nine features introduced in [56], [57] that can modify the elevator's behavior:

| | |
|---|---|
| Anti-prank | Normally, a button will remain pushed until the corresponding floor is served. With this feature, a button has to be held pushed by a person. |
| Empty | If the lift is empty, then all requests made in the cabin will be cancelled. |
| Executive floor | One floor has priority over others and is served first, both for cabin and for platform requests. |
| Open when idle | When idle, the lift opens its doors. |
| Overload | The lift will not close its doors when overloaded. |
| Park | When idle, the lift returns to the first floor. |
| Quick close | The doors cannot be kept open by holding the platform button pushed. |
| Shuttle | The lift will only change its direction at the first and the last floor. |
| Two-thirds full | Whenever the lift is two-thirds full, it will serve cabin calls before platform calls. |

The core logic of the controller is obviously subject to many constraints (e.g. the doors cannot be open while the elevator is moving) and the activated features add even more constraints (e.g. the Overload feature should impede to close the doors if the cabin is overloaded). Specifying all such constraints in an operational description is rather challenging and results in very sophisticated and cumbersome control flow statements (cf., e.g., [56]).

To show the effectiveness of QFLAN, we analyze a classical property of the Elevator SPL from [57] against variations obtained by increasing the number of floors from 5 to 40, while fixing the capacity of the elevator to 8 persons and the maximum allowed load to 4 persons (enforced only if the feature Overload is installed). Instead, the classical approaches mentioned above all restrict to models with less than 10 floors and fewer persons.

In particular, we analyze the property in Listing 13, which establishes that if the number of people in the elevator (the `load` variable) is beyond the capacity of the elevator (the `capacity` variable), then the elevator does not move (`direction == 0.0`). We check this property for all the states met in the first `maxStep` steps, i.e. until the condition `steps < maxStep` holds. In all cases, we obtained a probability equal to 1, because, by construction, the elevator does not move when the current load is higher than the capacity. Hence, all simulations performed in these analyses consisted of exactly `maxStep` steps.

```
begin analysis                                              1
  query = eval until { steps < maxStep } :                  2
              { load >= capacity implies direction == 0.0 } 3
  default delta = 0.1                                       4
  alpha = 0.1                                               5
  parallelism = 4                                           6
end analysis                                                7
```

Listing 13. Query to establish a safety property of the elevator

Figure 7 provides the runtimes (in seconds) of analyzing variants of the safety property for different values of `maxStep`, one per trace. In order to reduce stochastic noise, we provide runtimes averaged over 10 independent analyses. The figure provides two kinds of scalability analysis: (1) by focusing on a single trace we can fix the number of performed simulation steps, and vary the size of the system (the floors); (2) by considering one point of the $x$-axis (the floors), we can fix the model under analysis, and vary the number of performed simulation steps. In both cases, we note a linear increase in the obtained runtime.

All experiments were performed on a static pre-defined configuration consisting of all features except Park.

TABLE 4
Analysis effort in the Elevator with Configurator with varying number of reconfigurable features

| # Reconfigurable features | 0 | 1 | 2 | 3 | 4 | 5 | 6 | 7 | 8 | 9 |
|---|---|---|---|---|---|---|---|---|---|---|
| Analysis runtime (s) | 52.25 | 39.46 | 54.26 | 60.99 | 37.05 | 51.81 | 110.25 | 118.91 | 305.48 | 219.37 |
| # Simulations | 3,711 | 3,478 | 3,445 | 4,098 | 3,950 | 3,900 | 3,760 | 3,908 | 3,863 | 4,073 |
| Time per simulation (s) | 0.014 | 0.011 | 0.016 | 0.015 | 0.009 | 0.013 | 0.029 | 0.030 | 0.079 | 0.064 |

TABLE 5
Analysis effort in the Elevator with Configurator with varying number of features constrained by *requires* constraints

| # *requires* constraints | 0 | 1 | 2 | 3 | 4 | 5 | 6 | 7 | 8 |
|---|---|---|---|---|---|---|---|---|---|
| Analysis runtime (s) | 219.37 | 160.61 | 105.67 | 73.86 | 65.38 | 53.13 | 66.39 | 52.85 | 47.96 |
| # Simulations | 4,054 | 4,136 | 3,686 | 3,639 | 3,718 | 3,707 | 3,650 | 3,654 | 3,671 |
| Time per simulation (s) | 0.052 | 0.039 | 0.029 | 0.020 | 0.018 | 0.014 | 0.018 | 0.014 | 0.013 |

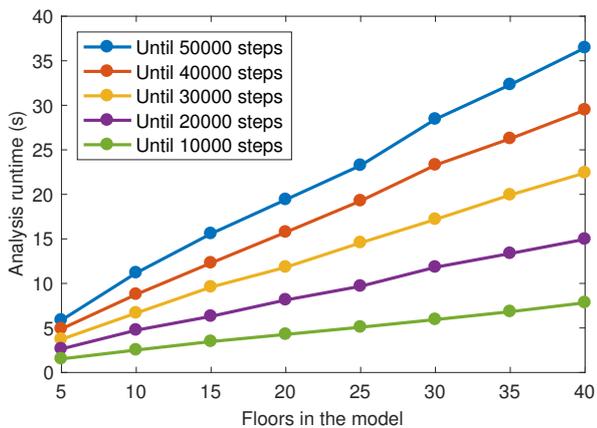

Fig. 7. Runtime in seconds for the analysis of Listing 13; each trace refers to variants of the property for `maxStep` in $\{10, 20, 30, 40, 50\} \times 10^3$

### 7.2 Elevator: Impact of reconfigurations

We present here an evaluation of the impact on analysis effort of one key characteristic of QFLan models, namely dynamicity of models in terms of reconfiguration frequency. Our initial hypothesis was that, the more dynamicity a model has, the more challenging the analysis would be since there may be more work for the interpreter (i.e. more actions and constraint checks to compute) and for the statistical analyzer (more variability should in principle require more simulations to achieve the same statistical confidence).

To validate our hypothesis, we have performed several times the same analysis on a set of variants of the Elevator SPL. In particular, we have enriched the model with a reconfigurator process which dynamically reconfigures the elevator controller by adding or removing features and we have parametrized such controller with the number of features that can be reconfigured, from 0 (none) to 9 (all).

The experiments we show here are based on an arbitrary order of the nine features but experiments with other orders yielded similar results. We have chosen a variant of the elevator with 4 floors, as this is already challenging for existing tools. For the analysis we have chosen a very demanding property, namely the floor where the elevator can be expected to be found on average, with a confidence interval of $\alpha = 0.1$ and $\delta = 0.1$ as in the previous section.

The results of the experiments are depicted in Table 4. We have measured the runtime (in seconds) and the number of simulations required. The table also includes the average runtime per simulation. The results in the table are the average of at least 10 different runs of the analyzer. This is necessary since simulations rely on random generators that are initialized with different seeds for every run so that the same analysis may actually produce different simulations and hence different runtime and simulations required.

The table shows that, indeed, the runtime tends to grow as the variability increases. The extreme case is a difference in runtime of about 7–8 times. Surprisingly, however, this is not always true. We have tried other variations of the model (e.g. changing the order of features) and of the analysis parameters (e.g. statistical confidence) and the obtained results are similar.

The variations in time do not seem to be directly related to the number of simulations needed. Indeed the number of simulations is quite stable and does not seem to be particularly influenced by the amount of features being reconfigured. Actually, the difference between the model that requires most simulations and the one that requires least simulations is around 10% only. We have further investigated this and we have discovered that some of the features, like `Parking` and in particular `Executive`, work as stabilizers of the system, as they attract the elevator to a particular floor.

In fact, the time needed per simulation witnesses that the variations in runtime are not mainly due to the number of simulations needed, but rather to the runtime effort needed to perform each single simulation. Indeed some simulations run about 8 times slower than others. We think that this is mainly due to the fact that the more features can be reconfigured the more transitions are potentially present for every state. The interpreter has hence more computations to do to actually compute those transitions and check their executability conditions (including consistency of the resulting state). However, this is not always the case, since some features disable actions rather than enabling them, which could explain why some models with more reconfigurable features take less time.

### 7.3 Elevator: Impact of feature constraints

This section presents an experimental evaluation similar to the one of Section 7.2. In particular, we evaluate the impact on analysis effort of another key characteristic of QFLAN models, namely product variability as imposed by feature constraints.

Our initial hypothesis was to observe a similar behavior as in the experiments of Section 7.2, i.e. that, up to some exceptions, less constrained models have more variability and require hence a greater analysis effort.

To validate our hypothesis, we have proceeded as in Section 7.2, i.e. we have performed several times the same analysis on a set of variants of the Elevator SPL with 4 floors. This time, we started with the model with the reconfigurator process described in Section 7.2 that dynamically reconfigures the elevator controller by adding or removing all nine features, and we have experimented by adding more and more *requires* constraints. In particular, we consider that all nine features are constrained by a chain of eight *requires* constraints and we have considered variations where such constraints are relaxed, up to the trivial case where there is no *requires* constraint among the features at all. The order of the features has been chosen arbitrarily, but other orders provided similar results. For the analysis we have chosen the same property and analysis parameters as in Section 7.2.

The results of the experiments are depicted in Table 5. The format of the table is the same as that of Table 4, i.e. we present runtime (in seconds), number of simulations and average runtime per simulation. Again, the results of the presented experiments show how there is an impact of the number of constrained features on analysis effort. In particular, the results show again that the main impact is not on the number of simulations but rather on the time needed to compute each simulation. Indeed, the simulations of the extreme case where all nine features are constrained by a chain of eight *requires* constraints run 4 times faster than those of the unconstrained model.

### 7.4 Safe lock

To illustrate the flexibility of our approach to model case studies from different application domains, we show in this section how QFLan can be used to perform risk assessment in security scenarios with high variability. In particular, we focus on the use of attack trees and the seminal example from that area, namely the Safe Lock [63].

Figure 8(a) presents the original attack tree from [63]. It provides a specification of a risk assessment for a safe lock system. An attack tree is essentially an and/or tree, where nodes represent goals, and sub-trees represent sub-goals. In this case, the root node represents the main threat being analyzed, namely the lock being opened by an attacker. Each of its four children are possible ways of enacting such a threat. The sub-goal Eavesdrop has two sub-goals that need to be accomplished (thus their combination as *and*-children). Nodes are decorated with an estimation of the cost that the attacker would have to pay to succeed in enacting the corresponding action. The classical analysis of such trees is to compute the minimal cost for an attacker to succeed.

Attack trees can easily be modelled as feature diagrams, with the following rationale: a node, which in an attack tree represents a goal, can be modeled as a feature of the system, that the attacker tries to activate. The sub-goal relation is modeled by the feature hierarchy. In particular, the attack tree of our case study can be modeled as in Fig. 8(b).

```
begin processes diagram                                      1
  begin process powerfulAttacker                             2
    states = idle                                            3
    transitions =                                            4
    idle -(install (PickLock) , 1)-> idle ,                 5
    idle -(install (CutOpenSafe) , 1)-> idle ,              6
    idle -(install (InstallImproperly) , 1)-> idle ,       7
    idle -(install (FindWrittenCombo) , 1)-> idle ,        8
    idle -(install (Threaten) , 1)-> idle ,                9
    idle -(install (Blackmail) , 1)-> idle ,               10
    idle -(install (ListenToConversation) , 1)-> idle ,    11
    idle -(install (GetTargetToStateCombo) , 1)-> idle ,   12
    idle -(install (Bribe) , 1)-> idle                     13
  end process                                               14
                                                            15
  begin process failingAttacker                             16
    states = idle , tryPickLock , tryCutOpenSafe ,         17
        tryInstallImproperly , tryFindWrittenCombo ,       18
        tryThreaten , tryBlackmail ,                       19
        tryListenToConversation ,                          20
        tryGetTargetToStateCombo , tryBribe                21
    transitions =                                           22
    // Try an attack                                        23
    idle -(try , 1 , { cumul_cost = cumul_cost + 1 })->    24
        tryPickLock ,
    idle -(try , 1 , { cumul_cost = cumul_cost + 1 })->    25
        tryCutOpenSafe ,
    idle -(try , 1 , { cumul_cost = cumul_cost + 1 })->    26
        tryInstallImproperly ,
    idle -(try , 1 , { cumul_cost = cumul_cost + 1 })->    27
        tryFindWrittenCombo ,
    idle -(try , 1 , { cumul_cost = cumul_cost + 1 })->    28
        tryThreaten ,
    idle -(try , 1 , { cumul_cost = cumul_cost + 1 })->    29
        tryBlackmail ,
    idle -(try , 1 , { cumul_cost = cumul_cost + 1 })->    30
        tryListenToConversation ,
    idle -(try , 1 , { cumul_cost = cumul_cost + 1 })->    31
        tryGetTargetToStateCombo ,
    idle -(try , 1 , { cumul_cost = cumul_cost + 1 })->    32
        tryBribe ,
    // Successful attack                                    33
    tryPickLock -(install (PickLock) , 1)-> idle ,         34
    tryCutOpenSafe -(install (CutOpenSafe) , 1)-> idle ,   35
    tryInstallImproperly -(install (InstallImproperly) , 1) 36
        -> idle ,
    tryFindWrittenCombo -(install (FindWrittenCombo) , 1)-> 37
        idle ,
    tryThreaten -(install (Threaten) , 1)-> idle ,         38
    tryBlackmail -(install (Blackmail) , 1)-> idle ,       39
    tryListenToConversation -(install (Listen-            40
        ToConversation) , 1)-> idle ,
    tryGetTargetToStateCombo -(install (GetTarget-         41
        ToStateCombo) , 1)-> idle ,
    tryBribe -(install (Bribe) , 1)-> idle ,              42
    // Failed attack                                        43
    tryPickLock -(fail , 10)-> idle ,                      44
    tryCutOpenSafe -(fail , 10)-> idle ,                   45
    tryInstallImproperly -(fail , 10)-> idle ,            46
    tryFindWrittenCombo -(fail , 10)-> idle ,             47
    tryThreaten -(fail , 10)-> idle ,                      48
    tryBlackmail -(fail , 10)-> idle ,                     49
    tryListenToConversation -(fail , 10)-> idle ,         50
    tryGetTargetToStateCombo -(fail , 10)-> idle ,        51
    tryBribe -(fail , 10)-> idle                           52
  end process                                               53
end processes diagram                                       54
```

Listing 14. Two kind of attackers for the safe lock scenario

We introduce a slight variation to overcome a well-known limitation of the original attack trees, namely the inability to encode the ordering of events. Indeed, Listen to Conversation should occur before Get Target to State Combo, which we can model with a *requires* cross-tree constraint. A feature model defines which configurations are valid, but not how (i.e. in which order) to configure them. QFLAN does model (re)configuration: features can be dynamically installed, removed or replaced as long as at any point in time all constraints are satisfied, including those imposed by the feature model. As noted in Section 3.8, the requires cross-tree constraint from Get Target to State

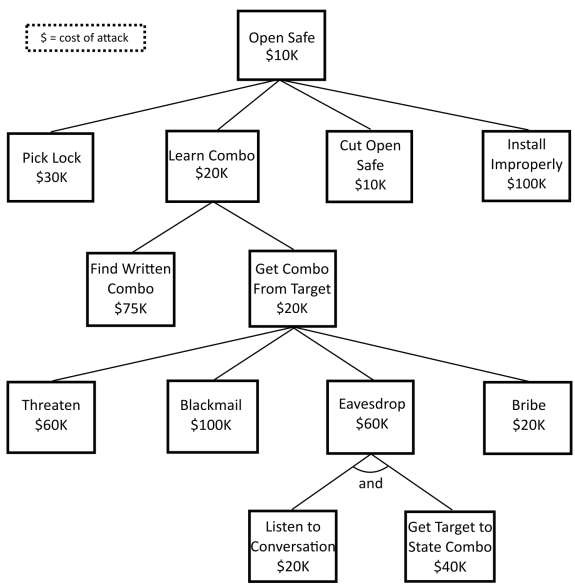

(a) Schneier's simple attack tree against a physical safe

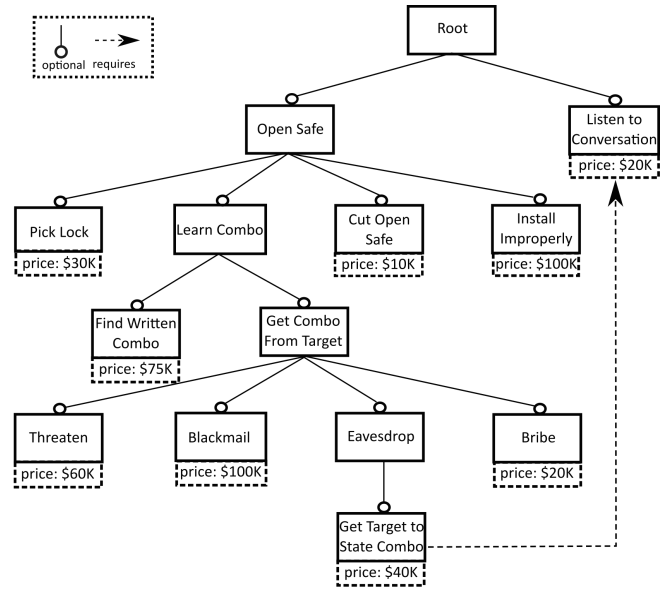

(b) An attributed feature model representing the attack tree

Fig. 8. Attack tree: (a) redrawn from [63]; (b) feature model representation of it

Combo to Listen to Conversation implies an order: whenever QFLAN tries to install (i.e. the attacker tries to activate) Get Target to State Combo, it fails to do so unless Listen to Conversation was installed (i.e. activated) before.

Hence, the flexibility of the way feature models are specified in QFLAN allows us to specify richer relations among sub-goals. For instance, we can specify that Eavesdrop is only successful if the attacker first listens to a conversation and then gets the target to state the combo, thus refining the original *and*-relation among such sub-goals. This is similar in spirit to the extension of attack trees with sequential conjunction from [64], which imposes orders on the execution of actions in the tree. Further constraints can be imposed on execution, in line with those discussed for the other examples.

A noteworthy advantage of modeling such scenarios with QFLAN is that we can model the behavior of several classes of attackers and study their performance, and thus the robustness of the system against them. To exemplify this, we have considered two attackers, sketched in Fig. 9. Their full process specifications can be found in Listing 14.

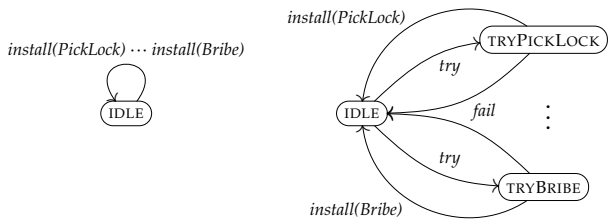

Fig. 9. Behavior of PowerfulAttacker (left) and FailingAttacker (right)

The Powerful attacker always succeeds when trying to achieve a goal and has unlimited resources. Instead, the Failing attacker can fail to successfully achieve a goal and may need several attempts to achieve them. This is modeled using rates. In addition, (s)he has limited resources.

Clearly, any reasonable attacker should stop attacking once an attack has been successful. This can be naturally expressed in QFLAN using the action constraints in Listing 15, which block the attacker after an attack succeeded.

```
begin action constraints                                    1
  do(tryAction) -> !has(OpenSafe)                           2
  do(install(...)) -> !has(OpenSafe)                        3
end action constraints                                      4
```

Listing 15. Constraints to stop attacks after success

QFLAN's rich specification language allows to express, e.g., further constraints on the accepted classes of attacks. Consider for instance the two constraints in Listing 16. In the first one, we restrict to (successful) attacks that cost less than $100K (i.e. that install features with less than that price, cf. the attributed feature model in Fig. 8(b)). Instead, the second constraint restricts to attacks (independently of their success) that cumulated less than 20 attempts. Noteworthy, using the first constraint we restrict the family of admissible products, while the latter constraint regards only the behavioral part of the model. In fact, `cumul_cost` is not an attribute but a variable, which can be changed through a memory update in the behavior (cf. Section 3.7). We use it as a counter to record the number of times that an attack is tried.

```
begin quantitative constraints                              1
  // Restrict to attacks that cost less than 100:           2
  { cost(Root) <= 100 }                                     3
  // Attacks can fail and attack attempts have a cost;      4
  // restrict to attackers that have a maximum budget:      5
  { cumul_cost <= 20 }                                      6
end quantitative constraints                                7
```

Listing 16. Further constraints to specify the class of accepted attacks

For both attackers, it is interesting to know what is the probability that an attack succeeds in a given amount of time, as well as the average cost of attacks. Such analyses can be performed in QFLAN, as shown in Listing 17. There we query the probability of installing the feature OpenSafe, the cost of the corresponding product and the attempts cumulated by the failing attacker while trying to install the features corresponding to the sub-goals. We consider three configurations: (a) a powerful attacker with constraints as specified above; (b) an attacker that might fail with the same constraints; and (c) an attacker that might fail with less resources, obtained by changing the constraint on the `cumul_cost` in Listing 16 to `{ cumul_cost <= 10 }`. This is obtained by running once the analysis on each model variant, each requiring about 12 seconds.

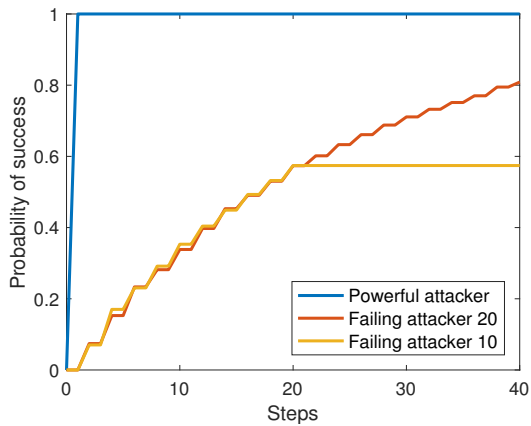

Fig. 10. Probabilities of successful attacks

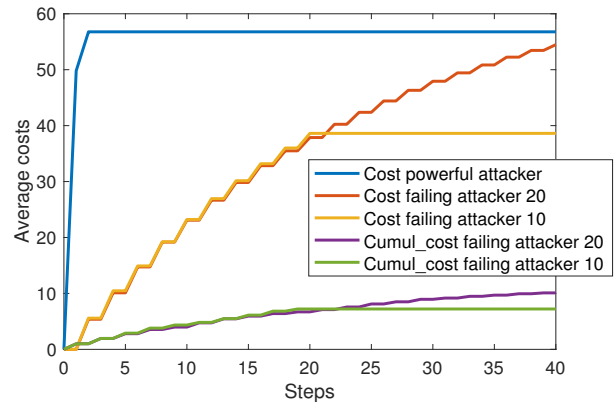

Fig. 11. Costs of successful attacks

```
1  begin analysis
2    query = eval from 0 to 40 by 1 :
3      { OpenSafe[delta = 0.05] , cost(Root) , cumul_cost }
4    default delta = 1
5    alpha = 0.05
6  end analysis
```

Listing 17. Analysis of the safe lock model

Figure 10 plots the probabilities of successful attacks. We note that the powerful attacker succeeds with probability almost 1 after one step, whereas for the other attacker the probability of success increases slowly. We also note that in case of constraint `{ cumul_cost <= 10 }`, the probability of success stabilizes at about 0.6 after 20 steps. Indeed, according to Listing 14, `cumul_cost` increases by 1 every two steps. Instead, in case `{ cumul_cost <= 20 }`, the probability reaches value 0.8 after 40 steps. However, this is not due to the mentioned constraint. In fact, Fig. 11 plots the costs and the cumulative attempts computed for the three model variants. We see that the average cumulative attempts (not shown for the powerful attacker, because it is always 0) reaches the threshold 10 after 20 steps (in case `{ cumul_cost <= 10 }`), while it is much lower than 20 in the other case. Hence, the constraint `{ cumul_cost <= 20 }` has less impact on the dynamics. As a last remark, we note that costs evolve similarly to probabilities, even though with different scales.

## 8  RELATED WORK

We concentrate on related approaches that focus on the application of automated verification techniques, and in particular (probabilistic) model checking, in the specific context of behavioral models of (dynamic) SPLs. We give a brief overview of models for specifying SPL behavior, followed by their associated verification techniques and tools.

### 8.1  Models

Most better known SPL behavioral modeling languages are based on superimposing multiple LTSs representing variants (products) in a single, enriched LTS (family) model. None of these languages allow the specification of probabilistic SPL models; the few languages that do will be discussed in Section 8.2 when we discuss probabilistic SPL model-checking approaches.

Featured Transition Systems (FTSs) were introduced in [53] and further elaborated in [59], [65]. An FTS models a family of LTSs (one per product) that can be obtained by projecting on the feature expressions (Boolean formulae defined over the set of features) decorating the transitions: all transitions whose feature expression is not satisfied by the specific product's set of features are removed, as well as all states and transitions that have become unreachable. Feature expressions are similar to action constraints in QFLAN but apply to transitions (i.e. instances of actions) rather than actions (i.e. classes of transitions). Feature expressions are more fine-grained, but action constraints provide a more compact and declarative specification. Adaptive FTSs, introduced in [65], allow the set of active features to vary dynamically, i.e. features can also be deactivated. QFLAN allows more general constraints than FTS' feature expressions (cf. Section 3.6) and adaptive or dynamic SPLs can be modeled as well; its action constraints are similar to the adaptation mechanisms of context-oriented programming as discussed and compared in [66].

Modal Transition Systems (MTSs) [67] were introduced to model successive refinements (implementations) of partial specifications. An MTS is an LTS distinguishing admissible (may) from necessary (must) transitions. In [68], MTSs were recognized as a suitable behavioral model for describing SPLs capable of checking the conformance of the behavior of a product against that of its product family. In a series of papers culminating in [10], MTSs were equipped with variability constraints which can express any Boolean function over the action labels, thus including all constraints that are typically defined by a variability model (but now expressed in terms of actions). Like FTSs, they model a family of LTSs (one per product) which can be obtained by turning each admissible but not necessary transition into a necessary transition or by removing it. Comparisons of FTSs and MTSs have appeared in the literature [52], [69]. Further MTS variants are variable I/O automata [70] and modal I/O automata [71]. QFLAN allows feature attributes and richer (quantitative) constraints (cf. Section 3.6) than any of these MTS models does, none of which moreover allows to model dynamic SPLs; the feature set is statically determined upfront.

Several process-algebraic theories for the modeling and analysis of SPLs have also been developed. In a series of papers, summarized in [72], Product Line CCS (PL-CCS) was defined as an extension of CCS by a variant operator allowing the user to model alternative behavior in the form of alternative processes, intending only one of them to exist at runtime. The choice calculus was introduced in [5] with the specific aim of providing a fundamental model for software variation, akin to the lambda calculus for programming languages. Another extension of CCS, DeltaCCS [11], was inspired by the well-known delta-modeling approach of automated product derivation for SPLs based on deltas that specify changes to be applied incrementally to a core product (cf. e.g. [73]). This modular approach differs from PL-CCS and the choice calculus, where choices are applied at well-defined variation points. Model-checking algorithms were implemented in MAUDE for SPLs specified in DeltaCCS against modal $\mu$-calculus formulas. In [9], a so-called Variant Process Algebra (VPA) is defined to formally reason on SPLs. Like [68], it focuses on behavioral (bi)simulation relations instead of verification through model checking. Our process-algebraic FLAN family distinguishes itself from all these approaches by the explicit modeling it offers for a rich set of constraints that may concern quantitative aspects of feature attributes. While PL-CCS and DeltaCCS allow some minimal restructuring functionality, none of these process-algebraic approaches can model dynamic SPLs.

Other known formalisms, quite different from QFLAN, that were equipped with variability notions concern Petri nets [55], [74], Event-B [75], Finite State Machines [76] and UML Activity Diagrams [77], [78]. In the latter, performance properties are captured by annotations (such as the duration to execute an activity of an activity node) and interpreted as CTMCs. Combined with the above mentioned delta-modeling approach, these constitute the first attempts to efficient performance modeling of SPLs.

### 8.2 Model checking

We now describe a number of SPL model-checking tools that have been introduced for the above modeling languages, followed by an overview of probabilistic SPL model-checking approaches. We are not aware of any other work than ours on statistical model checking for SPLs.

We only discuss approaches that, like ours, analyze behavioral models. There are also numerous SPL analysis approaches that operate directly on the source code, often obtained by adapting existing tools for software model checking to deal with variability. Examples include an adaptation of PROMOVER [79] with variability annotations [80] for Java and the SPLVERIFIER [81] tool chain built on JAVA PATHFINDER [82] for Java and CPACHECKER [83] for C code. SPLVERIFIER uses standard off-the-shelf model-checking techniques to verify the absence of feature interactions by an approach called feature-aware verification. For further details and for other model-checking approaches in SPLE than the ones described next, we refer to the survey [8] which covers not only SPL model checking, but also type checking, static analysis and theorem proving and which distinguishes *product-based*, *family-based* and *feature-based* analyses.

The analyses performed in this paper fall in the category of product-based analyses, according to which properties are verified on individually generated products (or at most a subset). Family-based analyses, on the contrary, concern the verification of properties on an entire product line, using variability knowledge about valid feature configurations to deduce results for individual products. Feature-based analyses, finally, concern the analysis of (domain artifacts implementing certain) features in isolation, which is not relevant to our approach given that features are only implicitly present as actions in QFLAN models.

With respect to typical product-based model checking, the statistical model-checking capabilities of QFLAN offer a couple of advantages. First, the set of simulations to be performed can be trivially parallelized and distributed over multiple cores, clusters or distributed computers with almost linear speedup. Second, the same set of simulations can be used to check several properties at the same time, thus requiring reduced computing time.

We first discuss several dedicated SPL model checkers.

The tool suite PROVELINES [84] supports discrete as well as real-time models, various types of computations, and advanced feature notions. All tool variants share the same common input language fPromela, which is an extension of the Promela input language of the well-known SPIN model checker (http://spinroot.com/). It includes the SPL model checker SNIP [85] for the verification of fLTL (feature LTL) properties over FTSs. A prototypical extension of the NuSMV model checker [86] uses a fully symbolic algorithm for the verification of fCTL (feature CTL) properties over FTSs specified in fSMV, which is a feature-oriented extension of the input language of (NU)SMV that was independently developed in the context of research on the renown problem of feature interaction [57]. In [23], SMT solving is implemented on top of SNIP (with Z3), i.e. behavioral models written in fPromela (with an FTS semantics) with additional arithmetic constraints. Admittedly, the resulting tool SNIP-Z3 does not scale well with the model size.

VMC (http://fmt.isti.cnr.it/vmc/) [87] is a tool for modeling and analyzing the behavior of SPLs modeled as MTSs with variability constraints [10]. Properties must be expressed in v-ACTL [88], a variability-aware action- and state-based branching-time temporal logic derived from the family of logics based on ACTL, the action-based version of CTL.

Next we discuss two off-the-shelf model checkers that were made amenable to SPL model checking, followed by a third one that offers probabilistic SPL model checking.

In [7], [89], a feature-oriented modular verification approach was developed, using an interpretation of FTSs in the MCRL2 formal specification language and toolset [90], [91]. MCRL2's parametrized data language allows one to handle feature attributes and quantitative constraints, like QFLAN. In order to perform family-based SPL model checking, a feature-oriented variant of the modal $\mu$-calculus, with an FTS semantics, was introduced in [92] by incorporating feature expressions into the modal operators, thus generalizing the work on the feature-oriented variants fLTL and fCTL. In [13], it was shown how to exploit this logic for family-based model checking with MCRL2 as-is by encoding it back into the logic of MCRL2.

In [93], [94], it was shown how to use SPIN for family-based model checking of LTL formulas against FTSs by means of an additional automatic variability-specific abstraction refinement method based on the discovery of spurious counterexamples obtained during model checking.

PROFEAT, a software tool built on top of PRISM for the analysis of feature-aware probabilistic models is presented in [34]. It provides a guarded-command language for modeling families of probabilistic systems as well as an automatic translation of family models to the input language of PRISM (i.e. featureless models). It can deal with probabilistic DSPLs by offering dynamic feature switching (i.e. activation and deactivation of features at runtime) and with feature attributes. The tool is evaluated through a number of case studies, including (probabilistic) versions of the Elevator benchmark SPL (cf. Section 7.1). Due to the nature of the analysis, i.e. statistical vs. precise probabilistic analysis, we were able to handle significantly larger variants of the Elevator SPL, viz. up to 40 floors rather than 4.

We close this section with a few pointers to probabilistic model checking of SPLs.

In [30], a Maple-based implementation is applied to a small running example (the usual coffee machine), while an empirical evaluation is limited to randomly generated behavioral models.

In [31], Discrete time Markov chain families (DTMCFs) are introduced as a model to specify the probabilistic behavior of an SPL. Moreover, a probabilistic model checking algorithm to verify Probablistic CTL (PCTL) formulas is defined. A tool is said to be forthcoming.

In [33], Featured DTMCs (FDTMCs) are introduced to model the probability of a transition being executed in a product. Verification of dependability (i.e. the probability to reach a success state) is formulated in PCTL. Furthermore, three family-based model-checking techniques are defined to verify stochastic SPLs modeled as FDTMCs derived from sequence diagrams.

In [32], Markov Decision Processes (MDPs) are used to model dynamic SPLs (in particular allowing the activation and deactivation of features at runtime), i.e. LTSs whose transitions have guards that formalize feature-dependent behavior annotated with probabilities and costs to model stochastic phenomena and resource constraints.

## 9 CONCLUSIONS AND FUTURE WORK

We have presented QFLAN, a quantitative modeling and verification environment for highly (re)configurable systems, such as dynamic SPLs, including Eclipse-based tool support. QFLAN offers a high-level DSL in which to specify system configurations and their probabilistic behavior as well as advanced statistical analyses of properties expressed in MultiQuaTEx. The QFLAN tool's GUI offers designers editing support typical of modern integrated development environments (such as auto-completion, syntax and error highlighting, etc.) for writing QFLAN specifications as well as MultiQuaTEx expressions.

We have shown a novel application of our approach to risk analysis of a safe lock system from the security domain as well as to classical examples of analysing the configuration and behavior of highly (re)configurable systems from the SPLE literature. Arguably the most important anomaly of feature models is the void feature model anomaly [24], i.e. when the root feature of the feature model cannot be selected thus forbidding the existence of any possible configuration. It is worth mentioning that QFLAN can actually be used to analyze the probability of this anomaly to occur, by verifying the probability of installing the root feature. The case studies have shown an analysis speedup for QFLAN, with respect to our earlier prototypical implementation, of more than three orders of magnitude.

We see a number of possible research directions for future work. First, we could provide additional QFLAN semantics for specific applications. For instance, a stochastic QFLAN semantics based on continuous time Markov chains to enable the analysis of time-related properties, an alternative QFLAN semantics based on FTSs to enable the use of the PROVELINES tool suite [84] or considering the general case of partially defined models.

Concerning the tool support, we could automatize its distributed analysis features (cf. Section 6), develop ad-hoc tool variants for attack trees (cf. Section 7.4), and improve interoperability with other tools, for instance for importing feature models designed with FeatureIDE [36], [37].

Finally, we could study the automatic synthesis of constraints starting from higher-level representations, possibly obtained by integrating our approach with the strategies synthesizer Uppaal Stratego [95].

## ACKNOWLEDGMENTS

Research supported by EU project QUANTICOL, 600708. We thank Bicincittà and M. Bertini from PisaMo for the bike-sharing case study.

## REFERENCES

[1] P. Clements and L. Northrop, *Software Product Lines: Practices and Patterns*. Addison-Wesley, 2002.

[2] K. Pohl, G. Böckle, and F. J. van der Linden, *Software Product Line Engineering: Foundations, Principles, and Techniques*. Springer, 2005.

[3] S. Apel, D. S. Batory, C. Kästner, and G. Saake, *Feature-Oriented Software Product Lines: Concepts and Implementation*. Springer, 2013.

[4] A. Gruler, M. Leucker, and K. D. Scheidemann, "Modeling and Model Checking Software Product Lines," in *Proceedings of the 10th International Conference on Formal Methods for Open Object-Based Distributed Systems (FMOODS'08)*, ser. LNCS, G. Barthe and F. S. de Boer, Eds., vol. 5051. Springer, 2008, pp. 113–131.

[5] M. Erwig and E. Walkingshaw, "The Choice Calculus: A Representation for Software Variation," *ACM Trans. Softw. Eng. Methodol.*, vol. 21, no. 1, 2011.

[6] A. Classen, M. Cordy, P.-Y. Schobbens, P. Heymans, A. Legay, and J.-F. Raskin, "Featured Transition Systems: Foundations for Verifying Variability-Intensive Systems and Their Application to LTL Model Checking," *IEEE Trans. Softw. Eng.*, vol. 39, no. 8, pp. 1069–1089, 2013.

[7] M. H. ter Beek and E. P. de Vink, "Using mCRL2 for the Analysis of Software Product Lines," in *Proceedings of the 2nd FME Workshop on Formal Methods in Software Engineering (FormaliSE@ICSE'14)*, S. Gnesi and N. Plat, Eds. ACM, 2014, pp. 31–37.

[8] T. Thüm, S. Apel, C. Kästner, I. Schaefer, and G. Saake, "A Classification and Survey of Analysis Strategies for Software Product Lines," *ACM Comput. Surv.*, vol. 47, no. 1, 2014.

[9] M. Tribastone, "Behavioral Relations in a Process Algebra for Variants," in *Proceedings of the 18th International Software Product Line Conference (SPLC'14)*, S. Gnesi, A. Fantechi, P. Heymans, J. Rubin, and K. Czarnecki, Eds. ACM, 2014, pp. 82–91.

[10] M. H. ter Beek, A. Fantechi, S. Gnesi, and F. Mazzanti, "Modelling and analysing variability in product families: Model checking of modal transition systems with variability constraints," *J. Log. Algebr. Meth. Program.*, vol. 85, no. 2, pp. 287–315, 2016.

[11] M. Lochau, S. Mennicke, H. Baller, and L. Ribbeck, "Incremental model checking of delta-oriented software product lines," *J. Log. Algebr. Meth. Program.*, vol. 85, no. 1, pp. 245–267, 2016.

[12] A. S. Dimovski, A. S. Al-Sibahi, C. Brabrand, and A. Wąsowski, "Efficient family-based model checking via variability abstractions," *Int. J. Softw. Tools Technol. Transf.*, pp. 1–19, 2016.

[13] M. H. ter Beek, E. P. de Vink, and T. A. C. Willemse, "Family-Based Model Checking with mCRL2," in *Proceedings of the 20th International Conference on Fundamental Approaches to Software Engineering (FASE'17)*, ser. LNCS, M. Huisman and J. Rubin, Eds., vol. 10202. Springer, 2017, pp. 387–405.

[14] M. H. ter Beek, A. Lluch Lafuente, and M. Petrocchi, "Combining Declarative and Procedural Views in the Specification and Analysis of Product Families," in *Proceedings of the 17th International Software Product Line Conference (SPLC'13)*, vol. 2. ACM, 2013, pp. 10–17.

[15] M. H. ter Beek, A. Legay, A. Lluch Lafuente, and A. Vandin, "Quantitative Analysis of Probabilistic Models of Software Product Lines with Statistical Model Checking," in *Proceedings of the 6th International Workshop on Formal Methods and Analysis for Software Product Line Engineering (FMSPLE'15)*, ser. EPTCS, J. M. Atlee and S. Gnesi, Eds., vol. 182, 2015, pp. 56–70.

[16] ——, "Statistical analysis of probabilistic models of software product lines with quantitative constraints," in *Proceedings of the 19th International Software Product Line Conference (SPLC'15)*, D. C. Schmidt, Ed. ACM, 2015, pp. 11–15.

[17] ——, "Statistical Model Checking for Product Lines," in *Proceedings of the 7th International Symposium on Leveraging Applications of Formal Methods, Verification and Validation: Foundational Techniques (ISoLA'16)*, ser. LNCS, T. Margaria and B. Steffen, Eds., vol. 9952. Springer, 2016, pp. 114–133.

[18] V. Saraswat and M. Rinard, "Concurrent Constraint Programming," in *Conference Record of the 17th Annual ACM Symposium on Principles of Programming Languages (POPL'90)*, F. E. Allen, Ed. ACM, 1990, pp. 232–245.

[19] M. G. Buscemi and U. Montanari, "CC-Pi: A Constraint-Based Language for Specifying Service Level Agreements," in *Proceedings of the 16th European Symposium on Programming (ESOP'07)*, ser. LNCS, R. De Nicola, Ed., vol. 4421. Springer, 2007, pp. 18–32.

[20] L. Bortolussi, "Stochastic Concurrent Constraint Programming," *ENTCS*, vol. 164, pp. 65–80, 2006.

[21] K. Czarnecki, S. Helsen, and U. W. Eisenecker, "Staged Configuration Using Feature Models," in *Proceedings of the 3rd International Software Product Lines Conference (SPLC'04)*, ser. LNCS, R. L. Nord, Ed., vol. 3154. Springer, 2004, pp. 266–283.

[22] J. Bürdek, S. Lity, M. Lochau, M. Berens, U. Goltz, and A. Schürr, "Staged Configuration of Dynamic Software Product Lines with Complex Binding Time Constraints," in *Proceedings of the 8th International Workshop on Variability Modelling of Software-intensive Systems (VaMoS'14)*, P. Collet, A. Wąsowski, and T. Weyer, Eds. ACM, 2014.

[23] M. Cordy, P.-Y. Schobbens, P. Heymans, and A. Legay, "Beyond Boolean Product-Line Model Checking: Dealing with Feature Attributes and Multi-features," in *Proceedings of the 35th International Conference on Software Engineering (ICSE'13)*. IEEE, 2013, pp. 472–481.

[24] D. Benavides, S. Segura, and A. Ruiz-Cortés, "Automated Analysis of Feature Models 20 Years Later: a Literature Review," *Inf. Syst.*, vol. 35, no. 6, 2010.

[25] J. Mauro, M. Nieke, C. Seidl, and I. C. Yu, "Context Aware Reconfiguration in Software Product Lines," in *Proceedings of the 10th International Workshop on Variability Modelling of Software-intensive Systems (VaMoS'16)*, I. Schaefer, V. Alves, and E. S. de Almeida, Eds. ACM, 2016, pp. 41–48.

[26] M. Nieke, G. Engel, and C. Seidl, "DarwinSPL: An Integrated Tool Suite for Modeling Evolving Context-aware Software Product Lines," in *Proceedings of the 11th International Workshop on Variability Modelling of Software-intensive Systems (VaMoS'17)*, M. H. ter Beek, N. Siegmund, and I. Schaefer, Eds. ACM, 2017, pp. 92–99.

[27] M. Clavel, F. Durán, S. Eker, P. Lincoln, N. Martí-Oliet, J. Meseguer, and C. Talcott, Eds., *All About Maude — A High-Performance Logical Framework: How to Specify, Program and Verify Systems in Rewriting Logic*, ser. LNCS, vol. 4350. Springer, 2007.

[28] S. Sebastio and A. Vandin, "MultiVeStA: Statistical Model Checking for Discrete Event Simulators," in *ValueTools*, A. Horvath, P. Buchholz, V. Cortellessa, L. Muscariello, and M. S. Squillante, Eds. ACM, 2013, pp. 310–315.

[29] L. M. de Moura and N. Bjørner, "Z3: An Efficient SMT Solver," in *Proceedings of the 14th International Conference on Tools and Algorithms for the Construction and Analysis of Systems (TACAS'08)*, ser. LNCS, C. R. Ramakrishnan and J. Rehof, Eds., vol. 4963. Springer, 2008, pp. 337–340.

[30] C. Ghezzi and A. Molzam Sharifloo, "Model-based verification of quantitative non-functional properties for software product lines," *Inform. Softw. Technol.*, vol. 55, no. 3, pp. 508–524, 2013.

[31] M. Varshosaz and R. Khosravi, "Discrete Time Markov Chain Families: Modeling and Verification of Probabilistic Software Product Lines," in *Proceedings of the 17th International Software Product Line Conference (SPLC'13)*, vol. 2. ACM, 2013, pp. 34–41.

[32] C. Dubslaff, C. Baier, and S. Klüppelholz, "Probabilistic Model Checking for Feature-Oriented Systems," in *Transactions on Aspect-Oriented Software Development XII*, ser. LNCS, S. Chiba, E. Tanter, E. Ernst, and R. Hirschfeld, Eds., vol. 8989. Springer, 2015, pp. 180–220.

[33] G. N. Rodrigues, V. Alves, V. Nunes, A. Lanna, M. Cordy, P.-Y. Schobbens, A. Molzam Sharifloo, and A. Legay, "Modeling and Verification for Probabilistic Properties in Software Product Lines," in *Proceedings of the 16th International Symposium on High-Assurance Systems Engineering (HASE'15)*. IEEE, 2015, pp. 173–180.

[34] P. Chrszon, C. Dubslaff, S. Klüppelholz, and C. Baier, "ProFeat: feature-oriented engineering for family-based probabilistic model checking," *Form. Asp. Comp.*, vol. 30, no. 1, pp. 45–75, 2018.

[35] M. Z. Kwiatkowska, G. Norman, and D. Parker, "PRISM 4.0: Verification of Probabilistic Real-Time Systems," in *CAV*, ser. LNCS, G. Gopalakrishnan and S. Qadeer, Eds., vol. 6806. Springer, 2011, pp. 585–591.

[36] T. Thüm, C. Kästner, F. Benduhn, J. Meinicke, G. Saake, and T. Leich, "FeatureIDE: An extensible framework for feature-oriented software development," *Sci. Comput. Program.*, vol. 79, pp. 70–85, 2014.

[37] J. Meinicke, T. Thüm, R. Schröter, F. Benduhn, T. Leich, and G. Saake, *Mastering Software Variability with FeatureIDE*. Springer, 2017.

[38] P. DeMaio, "Bike-sharing: History, Impacts, Models of Provision, and Future," *Journal of Public Transportation*, vol. 12, no. 4, pp. 41–56, 2009.

[39] P. Midgley, "Bicycle-Sharing Schemes: Enhancing Sustainable Mobility in Urban Areas," Background Paper CSD19/2011/BP8, Commission on Sustainable Development, United Nations Department of Economic and Social Affairs, May 2011.

[40] M. Bartoletti, T. Cimoli, M. Murgia, A. S. Podda, and L. Pompianu, "A Contract-Oriented Middleware," in *Proceedings of the 12th International Conference on Formal Aspects of Component Software (FACS'15)*, ser. LNCS, C. Braga and P. C. Ölveczky, Eds., vol. 9539. Springer, 2015.

[41] S. Arora, A. Rathor, and M. V. P. Rao, "Statistical Model Checking of Opportunistic Network Protocols," in *Proceedings of the Asian Internet Engineering Conference (AINTEC'15)*. ACM, 2015, pp. 62–68.

[42] L. Belzner, R. Hennicker, and M. Wirsing, "OnPlan: A Framework for Simulation-Based Online Planning," in *Proceedings of the 12th International Conference on Formal Aspects of Component Software (FACS'15)*, ser. LNCS, C. Braga and P. C. Ölveczky, Eds., vol. 9539. Springer, 2015, pp. 1–30.

[43] D. Pianini, S. Sebastio, and A. Vandin, "Distributed Statistical Analysis of Complex Systems Modeled Through a Chemical Metaphor," in *Proceedings of the International Conference on High Performance Computing & Simulation (HPCS'14)*. IEEE, 2014, pp. 416–423.

[44] S. Gilmore, M. Tribastone, and A. Vandin, "An Analysis Pathway for the Quantitative Evaluation of Public Transport Systems," in *Proceedings of the 11th International Conference on Integrated Formal Methods (IFM'14)*, ser. LNCS, E. Albert and E. Sekerinski, Eds., vol. 8739.  Springer, 2014, pp. 71–86.

[45] V. Ciancia, D. Latella, M. Massink, R. Paškauskas, and A. Vandin, "A Tool-Chain for Statistical Spatio-Temporal Model Checking of Bike Sharing Systems," in *Proceedings of the 7th International Symposium on Leveraging Applications of Formal Methods, Verification and Validation: Foundational Techniques (ISoLA'16)*, ser. LNCS, T. Margaria and B. Steffen, Eds., vol. 9952.  Springer, 2016, pp. 657–673.

[46] S. Sebastio, M. Amoretti, and A. Lluch Lafuente, "A Computational Field Framework for Collaborative Task Execution in Volunteer Clouds," in *Proceedings of the 9th International Symposium on Software Engineering for Adaptive and Self-Managing Systems (SEAMS'14)*, G. Engels and N. Bencomo, Eds.  ACM, 2014, pp. 105–114.

[47] L. Belzner, R. De Nicola, A. Vandin, and M. Wirsing, "Reasoning (on) Service Component Ensembles in Re- writing Logic," in *Specification, Algebra, and Software*, ser. LNCS, S. Iida, J. Meseguer, and K. Ogata, Eds., vol. 8373.  Springer, 2014, pp. 188–211.

[48] A. Legay, B. Delahaye, and S. Bensalem, "Statistical Model Checking: An Overview," in *Proceedings of the 1st International Conference on Runtime Verification (RV'10)*, ser. LNCS, H. Barringer, Y. Falcone, B. Finkbeiner, K. Havelund, I. Lee, G. J. Pace, G. Rosu, O. Sokolsky, and N. Tillmann, Eds., vol. 6418.  Springer, 2010, pp. 122–135.

[49] K. G. Larsen and A. Legay, "Statistical model checking: Past, present, and future," in *Proceedings of the 6th International Symposium on Leveraging Applications of Formal Methods, Verification and Validation (ISoLA'14)*, ser. LNCS, T. Margaria and B. Steffen, Eds., vol. 8802.  Springer, 2014, pp. 135–142.

[50] P. Asirelli, M. H. ter Beek, A. Fantechi, and S. Gnesi, "Formal Description of Variability in Product Families," in *Proceedings of the 15th International Software Product Lines Conference (SPLC'11)*, E. S. de Almeida, T. Kishi, C. Schwanninger, I. John, and K. Schmid, Eds.  IEEE, 2011, pp. 130–139.

[51] M. H. ter Beek and E. P. de Vink, "Software Product Line Analysis with mCRL2," in *Proceedings of the 18th International Software Product Line Conference (SPLC'14)*, vol. 2.  ACM, 2014, pp. 78–85.

[52] H. Beohar, M. Varshosaz, and M. R. Mousavi, "Basic behavioral models for software product lines: Expressiveness and testing preorders," *Sci. Comput. Program.*, vol. 123, pp. 42–60, 2016.

[53] A. Classen, P. Heymans, P.-Y. Schobbens, A. Legay, and J.-F. Raskin, "Model Checking Lots of Systems: Efficient Verification of Temporal Properties in Software Product Lines," in *Proceedings of the 32nd International Conference on Software Engineering (ICSE'10)*.  ACM, 2010, pp. 335–344.

[54] A. Fantechi and S. Gnesi, "Formal modeling for product families engineering," in *Proceedings of the 12th International Conference on Software Product Line Engineering (SPLC'08)*.  IEEE, 2008, pp. 193–202.

[55] R. Muschevici, J. Proença, and D. Clarke, "Feature Nets: behavioural modelling of software product lines," *Softw. Sys. Model.*, vol. 15, no. 4, pp. 1181–1206, 2016.

[56] A. Classen, M. Cordy, P. Heymans, A. Legay, and P.-Y. Schobbens, "Formal semantics, modular specification, and symbolic verification of product-line behaviour," *Sci. Comput. Program.*, vol. 80, no. B, pp. 416–439, 2014.

[57] M. Plath and M. Ryan, "Feature integration using a feature construct," *Sci. Comput. Program.*, vol. 41, no. 1, pp. 53–84, 2001.

[58] S. Apel, A. von Rhein, P. Wendler, A. Größlinger, and D. Beyer, "Strategies for Product-line Verification: Case Studies and Experiments," in *Proceedings of the 35th International Conference on Software Engineering (ICSE'13)*.  IEEE, 2013, pp. 482–491.

[59] A. Classen, P. Heymans, P.-Y. Schobbens, and A. Legay, "Symbolic model checking of software product lines," in *Proceedings of the 33rd International Conference on Software Engineering (ICSE'11)*.  ACM, 2011, pp. 321–330.

[60] N. Macedo, J. Brunel, D. Chemouil, A. Cunha, and D. Kuperberg, "Lightweight Specification and Analysis of Dynamic Systems with Rich Configurations," in *Proceedings of the 24th ACM SIGSOFT International Symposium on Foundations of Software Engineering (FSE'16)*.  ACM, 2016, pp. 373–383.

[61] J. Meinicke, C. Wong, C. Kästner, T. Thüm, and G. Saake, "On essential configuration complexity: measuring interactions in highly-configurable systems," in *Proceedings of the 31st IEEE/ACM International Conference on Automated Software Engineering (ASE'16)*, D. Lo, S. Apel, and S. Khurshid, Eds.  ACM, 2016, pp. 483–494.

[62] H. Sabouri, M. M. Jaghoori, F. S. de Boer, and R. Khosravi, "Scheduling and Analysis of Real-Time Software Families," in *Proceedings of the 36th Annual IEEE Computer Software and Applications Conference (COMPSAC'12)*, X. Bai, F. Belli, E. Bertino, C. K. Chang, A. Elçi, C. C. Seceleanu, H. Xie, and M. Zulkernine, Eds.  IEEE, 2012, pp. 680–689.

[63] B. Schneier, 1999. [Online]. Available: https://www.schneier.com/academic/archives/1999/12/attack_trees.html

[64] W. Lv and W. Li, "Space Based Information System Security Risk Evaluation Based on Improved Attack Trees," in *Proceedings of the 3rd International Conference on Multimedia Information Networking and Security (MINES'11)*.  IEEE, 2011, pp. 480–483.

[65] M. Cordy, A. Classen, P. Heymans, A. Legay, and P.-Y. Schobbens, "Model checking adaptive software with featured transition systems," in *Assurances for Self-Adaptive Systems: Principles, Models, and Techniques*, ser. LNCS, J. Cámara, R. de Lemos, C. Ghezzi, and A. Lopes, Eds.  Springer, 2013, vol. 7740, pp. 1–29.

[66] R. Bruni, A. Corradini, F. Gadducci, A. Lluch Lafuente, and A. Vandin, "A Conceptual Framework for Adaptation," in *Proceedings of the 15th International Conference on Fundamental Approaches to Software Engineering (FASE'12)*, ser. LNCS, J. de Lara and A. Zisman, Eds., vol. 7212.  Springer, 2012, pp. 240–254.

[67] K. G. Larsen and B. Thomsen, "A Modal Process Logic," in *Proceedings of the 3rd Symposium on Logic in Computer Science (LICS'88)*.  IEEE, 1988, pp. 203–210.

[68] D. Fischbein, S. Uchitel, and V. A. Braberman, "A foundation for behavioural conformance in software product line architectures," in *Proceedings of the ISSTA Workshop on Role of Software Architecture for Testing and Analysis (ROSATEA'06)*, R. M. Hierons and H. Muccini, Eds.  ACM, 2006, pp. 39–48.

[69] M. H. ter Beek, F. Damiani, S. Gnesi, F. Mazzanti, and L. Paolini, "From Featured Transition Systems to Modal Transition Systems with Variability Constraints," in *Proceedings of the 13th International Conference on Software Engineering and Formal Methods (SEFM'15)*, ser. LNCS, R. Calinescu and B. Rumpe, Eds., vol. 9276.  Springer, 2015, pp. 344–359.

[70] K. Lauenroth, K. Pohl, and S. Töhning, "Model Checking of Domain Artifacts in Product Line Engineering," in *Proceedings of the 24th International Conference on Automated Software Engineering (ASE'09)*.  IEEE, 2009, pp. 269–280.

[71] K. G. Larsen, U. Nyman, and A. Wasowski, "Modal I/O Automata for Interface and Product Line Theories," in *Proceedings of the 16th European Symposium on Programming (ESOP'07)*, ser. LNCS, R. De Nicola, Ed., vol. 4421.  Springer, 2007, pp. 64–79.

[72] M. Leucker and D. Thoma, "A Formal Approach to Software Product Families," in *Proceedings of the 5th International Symposium on Leveraging Applications of Formal Methods, Verification and Validation (ISoLA'12), Part I*, ser. LNCS, T. Margaria and B. Steffen, Eds., vol. 7609.  Springer, 2012, pp. 131–145.

[73] D. Clarke, M. Helvensteijn, and I. Schaefer, "Abstract Delta Modeling," *ACM SIGPLAN Not.*, vol. 46, no. 2, pp. 13–22, Oct 2010.

[74] H. Zhang, H. Zou, F. Yang, and R. Lin, "Modeling and Analysis of Behavioral Variability in Product Lines," *J. Inf. Comput. Sci*, vol. 9, no. 12, pp. 3589–3600, 2012.

[75] A. Gondal, M. Poppleton, and M. Butler, "Composing Event-B Specifications - Case-Study Experience," in *Proceedings of the 10th International Conference on Software Composition (SC'11)*, ser. LNCS, S. Apel and E. K. Jackson, Eds., vol. 6708.  Springer, 2011, pp. 100–115.

[76] J.-V. Millo, S. Ramesh, S. N. Krishna, and G. K. Narwane, "Compositional Verification of Software Product Lines," in *Proceedings of the 10th International Conference on Integrated Formal Methods (IFM'13)*, ser. LNCS, E. B. Johnsen and L. Petre, Eds., vol. 7940.  Springer, 2013, pp. 109–123.

[77] M. Kowal, I. Schaefer, and M. Tribastone, "Family-based performance analysis of variant-rich software systems," in *Proceedings of the 17th International Conference on Fundamental Approaches to Software Engineering (FASE'14)*, ser. LNCS, S. Gnesi and A. Rensink, Eds., vol. 8411.  Springer, 2014, pp. 94–108.

[78] M. Kowal, M. Tschaikowski, M. Tribastone, and I. Schaefer, "Scaling Size and Parameter Spaces in Variability-aware Software Performance Models," in *Proceedings of the 30th IEEE/ACM International Conference on Automated Software Engineering (ASE'15)*, ser. LNI, M. B. Cohen, L. Grunske, and M. Whalen, Eds., vol. 252.  IEEE, 2015, pp. 407–417.

[79] S. Soleimanifard, D. Gurov, and M. Huisman, "ProMoVer: Modular Verification of Temporal Safety Properties," in *Proceedings of the 9th International Conference on Software Engineering and Formal Methods (SEFM'11)*, ser. LNCS, G. Barthe, A. Pardo, and G. Schneider, Eds., vol. 7041. Springer, 2011, pp. 366–381.

[80] I. Schaefer, D. Gurov, and S. Soleimanifard, "Compositional Algorithmic Verification of Software Product Lines," in *Revised Papers of the 9th International Symposium on Formal Methods for Components and Objects (FMCO'10)*, ser. LNCS, B. K. Aichernig, F. S. de Boer, and M. M. Bonsangue, Eds., vol. 6957. Springer, 2012, pp. 184–203.

[81] S. Apel, H. Speidel, P. Wendler, A. von Rhein, and D. Beyer, "Detection of feature interactions using feature-aware verification," in *Proceedings of the 26th IEEE/ACM International Conference on Automated Software Engineering (ASE'11)*. IEEE, 2011, pp. 372–375.

[82] W. Visser, K. Havelund, G. P. Brat, S. Park, and F. Lerda, "Model Checking Programs," *Autom. Softw. Eng.*, vol. 10, no. 2, pp. 203–232, 2003.

[83] D. Beyer and M. E. Keremoglu, "CPAchecker: A Tool for Configurable Software Verification," in *Proceedings of the 23rd International Conference on Computer Aided Verification (CAV'11)*, ser. LNCS, G. Gopalakrishnan and S. Qadeer, Eds., vol. 6806. Springer, 2011, pp. 184–190.

[84] M. Cordy, A. Classen, P. Heymans, P.-Y. Schobbens, and A. Legay, "ProVeLines: a product line of verifiers for software product lines," in *Proceedings of the 17th International Software Product Line Conference (SPLC'13)*, vol. 2. ACM, 2013, pp. 141–146.

[85] A. Classen, M. Cordy, P. Heymans, A. Legay, and P.-Y. Schobbens, "Model checking software product lines with SNIP," *Int. J. Softw. Tools Technol. Transf.*, vol. 14, no. 5, pp. 589–612, 2012.

[86] A. Cimatti, E. M. Clarke, F. Giunchiglia, and M. Roveri, "NuSMV: A New Symbolic Model Verifier," in *Proceedings of the 11th International Conference on Computer Aided Verification (CAV'99)*, ser. LNCS, N. Halbwachs and D. A. Peled, Eds., vol. 1633. Springer, 1999, pp. 495–499.

[87] M. H. ter Beek, F. Mazzanti, and A. Sulova, "VMC: A Tool for Product Variability Analysis," in *Proceedings of the 18th International Symposium on Formal Methods (FM'12)*, ser. LNCS, D. Giannakopoulou and D. Méry, Eds., vol. 7436. Springer, 2012, pp. 450–454.

[88] M. H. ter Beek and F. Mazzanti, "VMC: Recent Advances and Challenges Ahead," in *Proceedings of the 18th International Software Product Line Conference (SPLC'14)*, vol. 2. ACM, 2014, pp. 70–77.

[89] M. H. ter Beek and E. P. de Vink, "Towards Modular Verification of Software Product Lines with mCRL2," in *Proceedings of the 6th International Symposium on Leveraging Applications of Formal Methods, Verification and Validation (ISoLA'14)*, ser. LNCS, T. Margaria and B. Steffen, Eds., vol. 8802. Springer, 2014, pp. 368–385.

[90] S. Cranen, J. F. Groote, J. J. A. Keiren, F. P. M. Stappers, E. P. de Vink, W. Wesselink, and T. A. C. Willemse, "An Overview of the mCRL2 Toolset and Its Recent Advances," in *Proceedings of the 19th International Conference on Tools and Algorithms for the Construction and Analysis of Systems (TACAS'13)*, ser. LNCS, N. Piterman and S. A. Smolka, Eds., vol. 7795. Springer, 2013, pp. 199–213.

[91] J. F. Groote and M. R. Mousavi, *Modeling and Analysis of Communicating Systems*. The MIT Press, 2014.

[92] M. H. ter Beek, E. P. de Vink, and T. A. C. Willemse, "Towards a Feature mu-Calculus Targeting SPL Verification," in *Proceedings of the 7th International Workshop on Formal Methods and Analysis for Software Product Line Engineering (FMSPLE'16)*, ser. EPTCS, J. Rubin and T. Thüm, Eds., vol. 206, 2016, pp. 61–75.

[93] A. S. Dimovski, A. S. Al-Sibahi, C. Brabrand, and A. Wąsowski, "Family-Based Model Checking Without a Family-Based Model Checker," in *Proceedings of the 22nd International SPIN Symposium on Model Checking of Software (SPIN'15)*, ser. LNCS, B. Fischer and J. Geldenhuys, Eds., vol. 9232. Springer, 2015, pp. 282–299.

[94] A. S. Dimovski and A. Wąsowski, "Variability-Specific Abstraction Refinement for Family-Based Model Checking," in *Proceedings of the 20th International Conference on Fundamental Approaches to Software Engineering (FASE'17)*, ser. LNCS, M. Huisman and J. Rubin, Eds., vol. 10202. Springer, 2017, pp. 406–423.

[95] A. David, P. G. Jensen, K. G. Larsen, M. Mikucionis, and J. H. Taankvist, "Uppaal Stratego," in *Proceedings of the 21st International Conference on Tools and Algorithms for the Construction and Analysis of Systems (TACAS'15)*, ser. LNCS, C. Baier and C. Tinelli, Eds., vol. 9035. Springer, 2015, pp. 206–211.